# LONG-LAG, WIDE-PULSE GAMMA-RAY BURSTS

J. P. Norris[1], J. T. Bonnell[1], D. Kazanas[1], J. D. Scargle[2],
J. Hakkila[3], and T. W. Giblin[3]

[1]Laboratory for High Energy Astrophysics
NASA/Goddard Space Flight Center, Greenbelt, MD 20771.

[2]Space Science Division
NASA/Ames Research Center, Moffett Field, CA 94035-1000.

[3]Department of Physics and Astronomy
College of Charleston, Charleston, SC 29424

ABSTRACT

Currently, the best available probe of the early phase of gamma-ray burst (GRB) jet attributes is the prompt gamma-ray emission, in which several intrinsic and extrinsic variables determine GRB pulse evolution. Bright, usually complex bursts have many narrow pulses that are difficult to model due to overlap. However, the relatively simple, long spectral lag, wide-pulse bursts may have simpler physics and are easier to model. In this work we analyze the temporal and spectral behavior of wide pulses in 24 long-lag bursts, using a pulse model with two shape parameters – width and asymmetry – and the Band spectral model with three shape parameters. We find that pulses in long-lag bursts are distinguished both temporally and spectrally from those in bright bursts: the pulses in long spectral lag bursts are few in number, and ~ 100 times wider (10s of seconds), have systematically lower peaks in $\nu F(\nu)$, harder low-energy spectra and softer high-energy spectra. We find that these five pulse descriptors are essentially uncorrelated for our long-lag sample, suggesting that at least ~ 5 parameters are needed to model burst temporal and spectral behavior. However, pulse width is strongly correlated with spectral lag; hence these two parameters may be viewed as mutual surrogates. We infer that accurate formulations for estimating GRB luminosity and total energy will depend on several gamma-ray attributes, at least for long-lag bursts. The prevalence of long-lag bursts near the BATSE trigger threshold, their predominantly low $\nu F(\nu)$ spectral peaks, and relatively steep upper power-law spectral indices indicate that Swift will detect many such bursts.

Subject headings: gamma rays: bursts, jets

# 1. LONG-LAG, WIDE-PULSE BURSTS

In the absence of gravitational wave and neutrino detections, the best probe of the early phase of gamma-ray burst (GRB) jet attributes is the prompt emission – essentially the temporal-spectral dependence of GRB pulses. It would appear that several jet and extrinsic parameters determine the pulse evolution: angular extent of emission cone; distance of emission region from central source; emitting shell thickness; profiles of Lorentz factor and mass density; evolution of emission spectrum with time and position, and view angle with respect to jet axis. However, some of these variables may be correlated, and it is not known a priori in any circumstance to what degree the number of independent observables is related to the number of physical determinants. Accurate study of individual pulse behavior is often difficult since in most bright bursts the pulses are bunched together and thus tend to overlap; in most dimmer bursts the difficulty is compounded by lower signal-to-noise ratios. In this circumstance, an expedient initial route to investigating the effective number of independent variables governing jet evolution is to study long-lag bursts, which tend to be dim, but also to have relatively simple temporal structure.

One definition of burst duration, $T_{90}$, is the time to accumulate between 5% and 95% of the total counts in a burst (Kouveliotou et al. 1993). Most bursts with $T_{90} > 2$ s detected by BATSE have many short, overlapping pulses (Norris, Scargle, & Bonnell 2001). But long-lag bursts have a few wide pulses with spectral lags (25–50 keV vs. 100–300 keV) of a few tenths to several seconds, and soft spectra (see Norris 2002, for details on measurement of spectral lag). Since their pulses are also long – a few to tens of seconds – the total number of detected photons per pulse can be comparable to that in bright bursts. With BATSE's DISCSC data type the pulse shapes are defined over many 64-ms time bins. Most important, since long-lag bursts tend to have just a few major pulses, the pulse confusion problem can be circumvented by careful sample selection, avoiding pulses which overlap substantially. However, it has also been shown that in at least two BATSE bursts, the predominant emission consists of pulses with short spectral lag, accompanied by a significantly delayed episode characterized by temporally smooth, low intensity, long-lag emission (Hakkila & Giblin 2004; Hakkila et al. 2004). The long-lag pulses in these two bursts have relatively high intensities compared to those of the short-lag pulses, incommensurate with expectations from the lag-luminosity relation (Norris, Marani & Bonnell 2000). How bursts with such disparate behavior may be related to bursts with exclusively long-lag emission – to be attributed to separate emission mechanisms or regions, or evolving jet properties – is part of the motivation for undertaking the present work.

Long-lag bursts appear to be important for other reasons as well. The proportion of these bursts increases from negligible among bright BATSE bursts to ~ 50% at trigger threshold (Norris 2002). An independent analysis using the burst "complexity parameter" previously revealed an admixture of a large fraction of simple bursts near BATSE trigger threshold (Stern, Poutanen, & Svensson 1999). The log [N] – log [$F_p$]] distribution for long-lag bursts follows a – 3/2 power-law over 1½ decades in peak flux ($F_p$). Bursts with very long lags (> 1–2 s) show a tendency (~ 95% confidence) to concentrate near the Supergalactic Plane, and thus may represent the low-luminosity tail of the GRB luminosity distribution. Only the nearest of such bursts could be detected by BATSE and would lie at distances of ~ 100 Mpc (Norris 2002). The one member clearly belonging to the long-lag group of bursts for which we have a redshift is GRB 980425 (SN 1998bw), with z = 0.0085, corresponding to a distance of 38 Mpc (Galama et al. 1998). Thus GRB 980425 would be consistent with the nearby hypothesis for long-lag bursts. This burst – and other low-redshift, low-luminosity bursts – may fit within the unified narrow jet model but observed far off axis (Lamb, Donaghy, & Graziani 2004), or may have a jet with wide collimation (Sazonov, Lutovinov & Sunyaev 2004; Soderberg et al. 2004).

The future observational consequences for a nearby subset of GRBs include possible detection in the gravitational wave and neutrino channels (Meszaros et al. 2003). The possibilities for study of long-lag bursts with Swift appear excellent since, as we show here, their peak in νF(ν) clusters near 110 keV, compared to the median value for bright BATSE bursts of ~ 230 keV (Preece et al. 2000), and their upper power-law indices (in the Band model context) are steep. Calculations by Band (2003) indicate that the sensitivity of the Burst Alert Telescope (BAT) on Swift to such bursts will be a factor of several better than BATSE's. This is attributable to the fact that the BAT effective area vs. energy curve has a broad peak in the range ~ 20 – 100 keV, decreasing to 50% of maximum area (mask-modulated) near ~ 15 and 150 keV (C. Markwardt, private communication).

A subclass of ultra-low luminosity, long-lag, soft-spectrum, nearby GRBs that could be produced by a version of the collapsar model was discussed by MacFadyen, Woosley, & Heger (2001). These would be the among the more common varieties of GRBs, but detected only locally. The explanations for their low luminosity include low Lorentz factors ($\Gamma$ ~ few: Kulkarni et al. 1998; Woosley & MacFadyen 1999; Salmonson 2001) compared to $\Gamma$ ~ $10^2$–$10^3$ for the high-luminosity bursts at cosmological distances; large viewing angle, profiled jets, or much wider jet opening angles. Kulkarni et al. and Wieringa, Kulkarni, & Frail (1999) inferred the latter for GRB 980425, based on the conclusion that the radio emission was not strongly beamed. The ultra-low luminosity of GRB 980425 may be attributable to membership on a steep second branch of the lag-luminosity relation for GRBs, related to mildly relativistic outflow (Kulkarni et al. 1998; Salmonson 2001; Norris 2002): if unlike the usual condition discussed for highly luminous GRBs, the initial ejecta have $\Gamma^{-1} > \theta_{jet}$, then different behavior is expected for the γ-ray and immediate afterglow phases. In fact, recently Guetta et al. (2004) conclude from a rate-density analysis for three low-luminosity bursts, which have the lowest redshifts (z: 0.0085, 0.105, and 0.17), that the luminosity function for GRBs is likely required to be a broken power-law in order to explain the occurrence of three such nearby bursts.

Thus there are central questions concerning both the observational contexts in which long-lag emission occurs and the physical conditions which give rise to these relatively simple bursts. Also, luminosity indicators have been suggested that include temporal and spectral measures (e.g., Norris, Marani & Bonnell 2000; Reichart et al. 2001; Amati et al. 2002). Since several physical variables combine to produce burst pulses, we search for any correlations between observed pulse attributes as a function of energy, attempting to constrain the minimum number of descriptors needed for long-lag bursts. These descriptors include pulse width and asymmetry as a function of energy, spectral lag, and spectral shape as expressed in the Band model. The details of our temporal and spectral analysis approach are described in section 2. The primary results are described in section 3, followed by a discussion in section 4. Characteristics of the pulse model are compiled in Appendix 1.

## 2. PULSE ANALYSIS METHODS

The sample of long-lag bursts studied here was culled from a set of 1429 BATSE bursts for which the spectral lags were measured near the BATSE trigger threshold (Norris 2002). The inclusion criterion for that original, larger sample and additional criteria for long-lag bursts are summarized in the next section. A four-parameter pulse model and its application to fitting simple burst temporal profiles are then described. The BATSE data types utilized for the pulse shape analysis are the 64-ms resolution PREB+DISCSC, which comprises four energy-loss channels: 25–50, 50–100, 100–300, and > 300 keV. The 1.024-s resolution DISCLA data was concatenated onto DISCSC data prior to PREB data for background fitting purposes. For spectral analysis, the MER and CONT data types with 16 energy-loss channels were utilized.

### 2.1 *Sample Selection*

Our long-lag burst sample inherits all the selection criteria described in Norris (2002) where spectral lags for 1429 bursts were measured. Briefly, bursts with $T_{90} > 2$ s, $F_p$ (50–300 keV) > 0.25 photons cm$^{-2}$ s$^{-1}$, and peak intensity (PI) > 1000 count s$^{-1}$ (> 25 keV) were included. Background fits and burst regions were defined, and peak fluxes and durations were measured following the same procedures described in Norris et al. (1996) and Bonnell et al. (1997). For apparently long-lag bursts, it was particularly necessary to examine carefully the 4-channel time profiles to ensure that a spurious lag was not measured: infrequently, sufficiently strong spectral evolution of the burst can result in, e.g., one pulse being most intense at high energy with the following pulse most intense at low energy. The resulting primary peak in the cross correlation then characterizes the interval between the two pulses, rather than the spectral evolution of individual pulses. Thus in the original study 12 bursts were eliminated in which spectral evolution of the overall burst, rather than pulse evolution, produced a spuriously long lag.

Several additional selection criteria yielded the long-lag sample analyzed here. The general objective was to include bursts with few pulses well defined over several to tens of seconds. The peak flux threshold was raised to $F_p > 0.75$ photons cm$^{-2}$ s$^{-1}$, and only 55 bursts with an average lag > 1 s were retained. These bursts were inspected visually in all four energy channels to assay if the pulses in each burst appeared sufficiently nonoverlapping to allow pulse fits with negligible ambiguity. Twenty-five were deemed too complex to proceed with a pulse fitting strategy, and in two bursts the pulses were too poorly defined. These 27 bursts were removed from further consideration. In some cases, bursts with one or two low-amplitude pulses which interfere negligibly with fitting the major pulses were tolerated. One special burst was added, BATSE trigger 6526 (GRB 971208). Even though an accurate background fit for this burst requires attention to multi-orbit phenomena (Connaughton 2002), we wished to obtain approximate pulse fits since it represents one extremum in single-pulse duration, as well as spectral lag, amongst BATSE bursts. Our peak flux and $T_{90}$ duration values for these 29 long-lag, few-pulse bursts with wide, well-defined major pulses are listed in Table 1. Brief comments are included concerning completeness of the MER data (used for the 16-channel spectral analysis) near the interval including the burst, and about the burst appearance. The latter is limited to a visual assessment of the number of major pulses, a remark if these pulses appear to overlap substantially, and the level of emission in channel 4 (> 300 keV). The qualifiers "very low, low, med, and high" for channel 4 indicate the following peak intensities, respectively: PI (counts s$^{-1}$) < 400, 400 < PI < 1000, 1000 < PI < 5000, and PI > 5000. The range in peak flux for bursts in Table 1 is 0.88 < $F_p$ (photons cm$^{-2}$ s$^{-1}$, 50–300 keV) < 3.58 (with one outlier at 15.33), and the spectral lag range is 1.0 < $\tau_{lag}$ (s) < 4.2 (with a different outlier at 30 s). Twelve of the 29 bursts have analyzable MER+CONT data – no gap occurs during the burst, or at most one energy channel of the 16 is absent. After performing the fitting procedures, we decided to exclude five more bursts from trend studies (BATSE triggers 1039, 2863, 6414, 6630, and 7969), due to discovered overlapping pulses (at least in some energy channels) which would have made interpretation ambiguous. We note that 11 bursts of our final 24 were part of a similar study by Kocevski, Ryde, & Liang (2003) aimed at bright bursts with relatively wide, separable pulses.

## 2.2 Temporal Analysis of BATSE LAD Data

The four channel DISCLA+PREB+DISCSC data were binned to 512 or 1024 ms resolution for further analysis. A Bayesian Block (BB) algorithm (Scargle et al. 2004) was run on the profile (> 25 keV) of each burst to find all significant pulse peaks and valleys between them, thus enumerating pulses destined to be fitted. Even though the algorithm works correctly with data binned to any timescale – finding any significant features which are not underresolved – we binned up in order to verify by visual inspection the results of the fitting procedures. BB representations obtained at the original 64-ms resolution did not reveal any significant temporal features at timescales shorter than the elected timescale for each burst.

From general experience with gamma-ray bursts we understand that a pulse model is desired which (1) minimizes unnecessary flexibility, namely a model with $\tau_{rise} < \tau_{decay}$, but otherwise allows for a continuum in asymmetry; (2) is continuously differentiable from pulse onset through peak and decay; and (3) reflects some general physical intuition regarding the source even though we are not treating the problem as a full-up radiation transfer. For fitting purposes it is also preferable that the partial derivatives with respect to the pulse shape parameters have relatively simple expressions. A form proportional to the inverse of the product of two exponentials, one increasing and one decreasing with time, satisfies the requirements:

$$I(t) = A \lambda / [\exp\{\tau_1/t\} \exp\{t/\tau_2\}] = A \lambda \exp\{-\tau_1/t - t/\tau_2\} \text{ for } t > 0, \quad (1)$$

where $\mu = (\tau_1/\tau_2)^{1/2}$ and $\lambda = \exp(2\mu)$. (The time of pulse onset with respect to t = 0, $t_s$, is ignored in eq. [1].) At $t = \tau_{peak} = (\tau_1\tau_2)^{1/2}$ the intensity is maximum, normalized by $\lambda$ to the peak intensity, $A$. The effects of $\tau_1$ and $\tau_2$ on the pulse amplitude are not completely intuitive – they are not respectively rise and decay constants – since their influence arises as the combined exponential in the denominator, with both terms operating across the pulse's duration. First notice that at pulse onset the first exponential alone drives the rise rate; however, since at pulse maximum, $\tau_1$ and $\tau_2$ contribute equally, $\tau_2$ already has asserted its effect during rise. After peak, the relative strength of the $\exp\{-\tau_1/t\}$ factor drops quickly, with the $\exp\{-t/\tau_2\}$ factor's dominance accelerating.

The pulse width measured between the two 1/e intensity points, is

$$w = \Delta\tau_{1/e} = \tau_2 (1 + 4\mu)^{1/2}. \quad (2)$$

Since $\mu \propto \tau_1^{1/2}$, $w$ is dominated by the $\tau_2$ timescale. At lower intensities, e.g., at $1/e^\nu$ intensity points ($\nu > 1$), $\tau_2$ dominance of the width is more accentuated, $\Delta\tau_{1/e^\nu} = \tau_2\nu(\nu + 4\mu)^{1/2}$. The form of the pulse asymmetry,

$$\kappa = (\tau_{decay} - \tau_{rise}) / (\tau_{decay} + \tau_{rise}) = \{(t_{1/e[decay]} - \tau_{peak}) - (\tau_{peak} - t_{1/e[rise]})\} / w, \text{ or} \quad (3)$$

$$\kappa = (1 + 4\mu)^{-1/2} = \tau_2 / w \quad (4)$$

has weak dependence on both $\tau_1$ and $\tau_2$, but allows a symmetric pulse in the limit as $\mu = (\tau_1/\tau_2)^{1/2} \to \infty$. From eqs. (2) and (4) we see that narrow, nearly symmetric pulses are produced for large $\tau_1$ and small $\tau_2$. For reference $\tau_{decay}$ and $\tau_{rise}$ are expressed in terms of $w$ and $\kappa$ as

$$\tau_{decay, rise} = \tfrac{1}{2} w (1 \pm \kappa) = \tfrac{1}{2} \tau_2 \{(1 + 4\mu)^{1/2} \pm 1\} \quad (5)$$

where the + (–) sign yields $\tau_{decay}$ ($\tau_{rise}$). Throughout our treatment we take the pair $w$ and $\kappa$ to be the quantities of interest, although $\tau_{decay, rise}$ would serve equally well with different emphasis.

The minimalist physical intuition of this model is that some (quasi) exponential process supplies energy (on timescale $\tau_1$) and another such process reduces a necessary condition for the continuance of the supply (on timescale $\tau_2$). For instance, the former could be a cascading injection of radiating particles and the latter, explosive expansion of the source. These combined exponential dependences in the model, while only phenomenological, turn out to afford good fits to the wide pulses in long-lag bursts. The partial derivatives necessary for implementing curve fitting and the expressions for error propagation for the pulse shape description are described in Appendix 1.

We note that the model expressed by eq. (1) provides pulse shapes that can be made very similar – by the adjustment of the $\tau_1$ and $\tau_2$ timescales – to the inverse of the difference of two exponentials, $[\exp\{\tau_1/t\} - \exp\{t/\tau_2\}]^{-1}$. However, the latter model has more complex partial derivatives. Alternatives such as variants on the inverse of the difference of power laws (Lazzati, Ramirez-Ruiz, & Ghisellini 2001; Schirber & Bullock 2003) were also implemented and exercised on the long-lag bursts, but sometimes the decay portions of these models produced extended tails and unacceptable fits.

Sufficient intervals were included before and after a burst to allow the pulse fit to descend to background level. Per energy channel, the BB algorithm was run to identify significant peaks and valleys, which points were utilized to calculate first guesses for the pulse shape parameter values. The chisquare-based fitting procedure usually converged to yield acceptable pulse fits, and overall burst fits, using this automated approach. However, manual intervention was required occasionally either to introduce a low-amplitude pulse, or to adjust the first guess values for a pulse. This was accomplished by visually estimating the peak, and 1/e rise and decay times and associated amplitudes for a pulse. See Appendix 1 for details on parameter value generation for the automated and manual intervention approaches. Table 2 lists pulse fit results per channel, including the fitted values for the parameters of eq. (1) as well as the derived parameters; these results are discussed in detail in section 3.

### 2.3 *Spectral Analysis of BATSE MER Data*

From our sample of 24 bursts for which we obtained useful fits of the pulse shapes, twelve had complete MER (16-channel) coverage of the burst interval, or else any gap present could be tolerated. The exceptions with complete MER data were trigger 764, which was not available at the time of the analysis; and trigger 6526 for which we could not fit an accurate background since the burst spanned more than one orbit (Connaughton 2002). The twelve bursts for which we performed spectral fits are indicated in Table 1. Each region to be analyzed included that portion of a major pulse where the intensity is above $0.2 \times$ the peak intensity. In eleven cases this region was divided into two intervals for spectral analysis with equal counts above background (> 25 keV) in the two intervals. The pulse division occurred near the peak for the energy range ~ 25–100 keV, after the peak at higher energies. For trigger 3256 only one interval was used since the pulse fluence was relatively low. One major pulse per burst was fitted except in the case of trigger 8049 where two major pulses were treated. Sometimes a low-intensity (essentially negligible) overlapping pulse intruded into a spectral fit interval.

For fitting the gamma-ray burst spectra we used the usual Band model (Band et al. 1993) forward folded through a burst-specific detector response matrix to directly fit the counts spectrum. The Band model effects a smooth join between two power laws:

$$I(E) = A\,[E/E_{piv}]^{\alpha}\,\exp(-E/E_0) \qquad\qquad E \leq (\alpha - \beta)E_0$$
$$\phantom{I(E)} = A\,[(\alpha - \beta)E_0/E_{piv}]^{(\alpha - \beta)}\,\exp(\beta-\alpha)\,[E/E_{piv}]^{\beta} \qquad E > (\alpha - \beta)E_0 \qquad (6)$$

where $E_{piv} = 100$ keV. We did not use channels 1 (~ 16–25 keV) and 16 (> 1850 keV) in the spectral fits. Channel 1 sometimes does not comply with the lower power-law of eq. (6); channel 15 is open-ended. The initial parameter values were $A = 1$, $\alpha = -0.5$, $\beta = -2.5$, and $E_0 = 100$ keV. Manual intervention was infrequently required to guide the fitting program towards an acceptable fit. The resulting fit parameters of interest for investigation of possible correlations with pulse fit parameters are: $\alpha$, $\beta$, and the peak in the $\nu f(\nu)$ spectrum, $E_{pk} = (2+\alpha)E_0$. Table 3 lists the fit intervals and the values for these spectral parameters.

We note that forms analogous to eq. (1) containing the inverse of the sum of power laws, which have a single, simpler form than eq. (6), may be used to fit smoothly joined power laws, but we have not explored the possibility (e.g., see eq. [1] of Schirber & Bullock 2003).

3. RESULTS

To recapitulate, temporal fits using the model expressed in eq. (1) were performed for the 24 long-lag bursts in which major pulses were sufficiently well defined and any overlap with minor pulses was tolerable. Table 2 lists the fitted parameter values for all identified pulses, but only those pulses with peak intensities greater than 0.25 times that of the most intense pulse were retained for illustration in the plots described below. The parameters included in Table 2 are: the burst identifier (BATSE trigger number), channel number, reduced chi-squared, pulse peak intensity ($A$), pulse onset time ($t_s$), "effective" onset time ($t_{eff}$), peak time ($\tau_{peak}$), the two fundamental timescales ($\tau_1$ and $\tau_2$), and the derived shape parameters, width ($w$) and asymmetry ($\kappa$). The associated error is listed below each parameter value. The formal time of pulse onset, $t_s$, often occurs at an intensity several orders of magnitude below the peak intensity, in which case $t_s$ is not indicative of the visually apparent onset time, $t_{eff}$, arbitrarily defined here as the time when the pulse reaches 0.01 times the peak intensity. Both onset times are relative to burst trigger time, which is listed in the BATSE online catalog (need ref). The listed peak time is calculated relative to the effective onset time. Illustrative plots of some representative burst fits from Table 2 are shown in Appendix 2.

3.1 *Temporal Shape*

Figures 1 through 8 illustrate correlations, or lack thereof, for retained pulses with peak intensities greater than ¼ that of the most intense pulse within a burst. For most bursts, this criterion results in only one pulse surviving for inclusion in these scatter plots. Note that the number of retained pulses varies with energy channel.

Figure 1 illustrates scatter plots for the two fundamental shape parameters, $\tau_2$ vs. $\tau_1$, for three energy channels. These parameters are fundamental, only in the sense that they are two of the four fitted parameters, not in the physical sense. Recall that their effects in eq. (1) operate across the duration of the pulse, with the $\exp\{-\tau_1/t\}$ factor dominating at onset, comparable effects near the peak time, $\tau_{peak} = (\tau_1\tau_2)^{1/2}$, and the $\exp\{-t/\tau_2\}$ factor prevailing during the decay. In assessing the possibility of a trend, we will usually have the luxury of three nearly statistically independent energy channels (in channel 4 the pulse is usually of too low intensity to be dependably useful, or else it is absent). Hence we should require clear indication of a trend in at least two adjacent channels. In Figure 1a (channel 1) we see a possible anticorrelation – with large dispersion – between $\log(\tau_1)$ and $\log(\tau_2)$. The dashed line indicates the best linear fit to this tentative trend, $\log(\tau_2) = -0.28 \times \log(\tau_1) + 1.22$, or $\tau_2 \approx 16.6 \times \tau_1^{-0.28}$. In Figures 1a and 1b (channels 2 and 3) the sense of this trend is the same, but with closer to zero slope, the trend marginally suggestive ($\tau_2 \propto \tau_1^{-0.21}$ and $\tau_2 \propto \tau_1^{-0.08}$, respectively). Note that Figure 1 and the trend fits exclude results for trigger 6526 – which has $\tau_1 \sim 20$ and $\tau_2 \sim 250$ nearly independent of energy – also an outlier in all other scatter plots discussed below.

Generally we should ask which channel(s) to place more emphasis on, given that the set of pulse fits for each channel are essentially equal in terms of goodness of fit ($\chi^2/\nu$). Pulses are invariably longer at lower energies, hence the pulses in channel 1 are always defined over longer intervals. However, the pulses in channels 2 and 3 usually have more counts and therefore higher quality definition. The suggestive anticorrelation in Figure 1 is less clear and has wider dispersion as energy increases. We could conclude that the anticorrelation is either a function of energy, or that it is spurious. However, the picture appears clearer when we consider the two derived pulse shape parameters.

Figure 2 illustrates scatter plots for asymmetry and width. For all three energy channels the centroid in asymmetry is $\sim 0.45$, with ranges $0.1 \lesssim \kappa \lesssim 0.7$, the distribution tightening slightly to higher energy. The average asymmetry combining all four energy channels is $\sim 0.45$ with a full-width at half maximum of $\sim 0.10$. This is only a somewhat broader distribution, with a similar mean, than reported for a sample of separable pulses in BATSE bursts by Kocevski, Ryde, & Liang (2003), who find $\kappa = 0.47 \pm 0.09$. The similarity argues that single-pulse bursts and wide pulses in GRBs in general, are not distinguishable based on pulse asymmetry. Recall that asymmetry and width form a pair with information equivalent to the pair $\tau_{decay}$ and $\tau_{rise}$. Asymmetries of 0.1, 0.4, and 0.7 correspond to rise-to-decay ratios, $\{1 - \kappa\}/\{1 + \kappa\}$, of 0.82, 0.43, and 0.18, respectively. The pulse width manifests an order of magnitude dispersion independent of energy, with the centroid shifting from 20 s to 10 s as energy channel increases. Again the one obvious outlier in width (300–400 s) is trigger 6526 – the longest single pulse burst – with $\kappa \sim 0.67$ independent of energy.

There appears to be no trend of width with asymmetry. But we might expect a correlation since $w/\kappa = \tau_2 (1 + 4\mu)^{1/2}$. Hence $w/\kappa$ is approximately proportional to $(\tau_1\tau_2)^{1/2}$ for $\tau_1/\tau_2 > 1/2$ (80% of fitted pulses) and increasing $\tau_2$ or $\tau_1$ would increase the ratio $w/\kappa$ ($w/\kappa$ is nearly proportional to $\tau_2$ for $\mu < 1$, 20% of pulses). But the anticorrelation of $\tau_2$ and $\tau_1$ and the specific distributions of these parameters will tend to cancel the expected correlation: Table 4 lists the $\pm 1\,\sigma$ ranges for $\tau_1$, $\tau_2$, $\mu$, and the co-factor for $w/\kappa$, $(1 + 4\mu)$. In fact, notice that for each channel the ranges and directions of variation for the cofactors of $w/\kappa$ approximately cancel each other. Moreover, the dynamic ranges for $\tau_1$ are 2–2½ orders of magnitude compared to those for $\tau_2$ of 1–1½ orders of magnitude, both ranges decreasing with energy. So the larger dispersion in $\tau_1$ contributes more to the total dispersion seen in the $\kappa$–$w$ diagrams of Figure 2. The dynamic ranges of $\tau_2$ with $\tau_1$ and their slight anti-correlations are such that the selected quantities of interest, pulse width and asymmetry, are essentially uncorrelated in our long-lag burst sample.

A similar appraisal of $\tau_2$ and $\tau_1$ dependences can be made for the asymmetry versus peak time plots, shown in Figure 3. For the large majority of pulses, $\kappa$ slowly decreases as $(\tau_1/\tau_2)^{-1/4}$. $\tau_{peak}$ increases as $\tau_1^{1/2}$ and $\tau_2^{1/2}$; therefore $\kappa$ and $\tau_{peak}$ would tend to be (anti)correlated due to the effect of $\tau_2$ ($\tau_1$). Since $\tau_1$ has a larger dynamic range than $\tau_2$ by a factor of ~ 3–10, the a priori combined effect (if $\tau_1$ were completely uncorrelated with $\tau_2$) would be an anticorrelation between $\kappa$ and $\tau_{peak}$ with large dispersion. The slight anticorrelation of the actual distributions of $\tau_1$ and $\tau_2$ evident in Figure 1a (possibly in Figures 1b and 1c) has the effect when translated to the $\tau_{peak}$–$\kappa$ plane of compressing (weakly expanding) the $\tau_{peak}$ ($\kappa$) dynamic range. In fact, the expected anticorrelation is apparent, with considerable dispersion, in all three parts of Figure 3. The fitted relations – $\kappa = p \times \log(\tau_{peak}) + \kappa_0$, or $\tau_{peak} = \exp[(\kappa - \kappa_0)/p]$ – are listed in Table 5. Similar slopes obtain across the three energy channels, flattening from 0.42 to 0.30 with increasing energy channel. In summary, for $\kappa$ and $\tau_{peak}$ an anticorrelation is "built in" by the form of eq. (1), and is not nullified by the actual distributions of $\tau_1$ and $\tau_2$ in Figure 1 as was the case for width and asymmetry. Thus $\kappa$ and $\tau_{peak}$ are not independent, by initial design.

Similarly, the dependence of $w$ on $\tau_2$ and $\tau_1$ is approximately $\propto \tau_1^{1/4} \tau_2^{3/4}$ for the large majority of pulses ($\tau_1/\tau_2 > \frac{1}{2}$). So with $\tau_{peak} \propto (\tau_1 \tau_2)^{1/2}$ a strong correlation is expected, again by design. This is apparent in Figure 4 for all three channels. The slope would be predicted to be near unity if $\tau_1$ and $\tau_2$ were uncorrelated. However, the $\tau_1$–$\tau_2$ anticorrelations of Figure 1 reduce the expected slopes, less so as energy channel increases. The fitted parameters for the power laws plotted in Figure 4 and listed in Table 6 confirm the expectation, with the slope increasing from 0.7 to 1.0 as energy channel increases. Also note that the $w$–$\tau_{peak}$ correlation is tightest in channel 3. Since $\tau_{peak}$ is a measure of the pulse rise time, the conclusion is that as pulse width increases, the rising portion of a pulse is longer, and slightly more so at higher energies: wider pulses tend to be slightly closer to symmetric at higher energy. Again, the points for trigger 6526 were not used in the fits for Figures 3 and 4. Even though the fitted trends in Figure 4 would appear to be continued to wider pulses by inclusion of trigger 6526, due to its outlying positions in the $w$–$\kappa$ and $\tau_1$–$\tau_2$ planes we elected not to bias the fitted $w$–$\tau_{peak}$ relations with pulses which are an order of magnitude wider (in each channel) than the next widest pulse in the sample.

From the start we understood that the three parameters – $w$, $\kappa$ and $\tau_{peak}$ – have exactly two degrees of freedom between them. Taking $w$ and $\kappa$ as the fundamental parameters of interest, the primary knowledge that we gain from the treatment so far is the lack of correlation between $w$ and κ and thus their apparent independence. The correlations in the $w$–$\tau_{peak}$ and $\kappa$–$\tau_{peak}$ planes are expected from eq. (1), and are only modified in degree by the correlations between $\tau_1$ and $\tau_2$, and their differing dynamic ranges, appreciated in Figure 1 and Table 4.

One of our two chosen fundamental temporal variables, pulse width, is of course a defining feature of the long-lag sample. Figures 5 and 6 emphasize the stark difference between pulse widths in long-lag bursts and those that are prevalent in bright bursts. Figure 5 illustrates individual pulse widths fitted for each energy channel; the corresponding width histograms are shown in Figure 6. The mode shifts from 20–30 s (25–50 keV) to 6–10 s (> 300 keV). The same general trend is evident as in all bursts – pulses narrow at higher energies (for the three discrepant cases in Figure 5 the 1-σ errors are larger than the difference in widths in the adjacent channels). But long-lag pulses have widths of several to tens of seconds, *two orders of magnitude longer* than pulses in the brightest long bursts, with widths of ~ 1–few × 100 ms. Yet the two different samples have similar average power-law dependences with energy. Fitting the average width (in the logarithm) per channel as a function of the geometric means of the lower and upper channel boundaries (using 300–1000 keV for channel 4) yields the representation (trigger 6526 excluded),

$$w\,(s) \approx 85.5\,E^{-0.41}\,. \qquad (7)$$

The power law representation for the widths of much narrower (separable) pulses in bright bursts was found to be nearly identical to eq. (7) by Fenimore et al. (1995), who used an autocorrelation approach for measuring pulse widths; and represented by a $w \propto E^{-0.35}$ dependence with a sample error of ~ 10%, as reported in Norris et al. (1996). The similar power-law dependences for pulse width provides evidence that the governing physics is related for the two extremes of the GRB width and peak flux distributions.

As pulse width increases, the spectral lag measured between pulse peaks tends to increase, as shown in Figure 7. The lags plotted are between channels 2 and 3; pulse width is for channel 2. The relative error sizes are large for peak lag; other channel combinations have even larger errors. The reason for this is evident from eq. (A9): the contributions to the peak lag error budget include the errors for both peaks and both pulse start times. The former components dominant due to the roundedness of pulse peaks.

Lag of pulse centroid, for which start time ($t_{eff}$) dominates the error budget (eq. A10), is much better measured as is illustrated in Figure 8. Centroid lag is well correlated with pulse width in Figure 8a where $\tau_{lag13}$ is between channels 1 and 3 (fewer pulses exist for $\tau_{lag24}$, Figure 8b). The power-law fit over one decade in pulse width shown in Figure 8a yields

$$\tau_{lag13} \text{ (s)} \approx 0.089 \, w^{1.42} \tag{8}$$

with a Spearman rank coefficient of $\sim 0.8$ ($P \sim 4\times10^{-6}$). Thus centroid lag may provide a surrogate measurement of pulse width, or vice versa. Width is better measured for individual pulses that are effectively separated, as is often the case for long-lag bursts. However, the numerous narrower pulses in brighter bursts frequently overlap (Norris et al. 1996), making measurement of pulse width difficult, whereas calculation of the average spectral lag in a crowded pulse intervals is straightforward. In the proposed relationship between spectral lag and peak luminosity for relatively bright bursts (Norris, Marani, and Bonnell 2000; Norris 2002), pulse width could be a more fundamental quantity than spectral lag, but it is often virtually impossible to measure for individual pulses.

In summary: The pulse shape expressed in eq. (1) is defined by just two of the four fitted parameters – $\tau_1$ and $\tau_2$ (Figure 1) – which can be transformed into a pair of pulse quantities of frequent inquiry – width (eq. 2) and asymmetry (eq. 3). This pair is taken as our primary pulse shape parameters of interest. Pulse asymmetry and width are found to be essentially uncorrelated (Figure 2) in channels 1–3, indicating that at least two physical variables operate to determine pulse shape. The time of pulse peak ($\tau_{peak}$, a measure of rise time with respect to effective pulse onset time) is anticorrelated with asymmetry (Figure 3), and well correlated with width (Figure 4), but these correlations are in fact expected from the form of eq. (1). In our sample of long-lag bursts, the pulse widths range from several to tens of seconds (Figures 5 and 6), two orders of magnitude wider than the numerous narrow pulses that predominate in bright bursts. However, the power-law variation of pulse width with energy (eq. 7) has a similar scaling to that found for narrow pulses in bright bursts. Finally, peak lag and centroid lag, the better measured quantity of the two, are correlated with width (Figures 7 and 8). One of the motivations for this study was to help elucidate the apparent dependence of luminosity on spectral lag. Hence a central question to be addressed for the relation between lag and width (eq. 8) is its extension to shorter pulses in bright bursts.

### 3.2 *Spectral Shape*

Table 3 lists the values for the spectral parameters of the Band model – $\alpha$, $\beta$, and $E_{pk}$ – for the 12 of 24 bursts where spectral analysis was possible. Usually, two spectral intervals were specified per pulse, with the division occurring near the time of pulse maximum ($\tau_{peak}$) for the energy range of channels 1+2. Except in the case of trigger 8049, the burst contained only one major pulse. The results from all spectral fits are included in Figures 9–11 and Figures 14–22. Figures 12 and 13 compare the hardness ratios for dim bursts with short and long lags.

Figures 9–11 show the distributions for the fitted spectral parameters. All three distributions have significantly different modes than those of bright burst sample analyzed by Preece et al. (2000). Note that multiple spectral intervals per burst were fitted in their treatment, as in ours. For their sample the mode of the lower power-law index is $\alpha_{brt} \approx -0.95$, with full width at half maximum (FWHM) of $\Delta\alpha_{brt} \approx 0.8$; the upper power-law index mode is $\beta_{brt} \approx -2.25$, tailing off by $\beta_{brt} \sim -4$ with higher negative indices up to $\beta_{brt} \sim -7$, and FWHM of $\Delta\beta_{brt} \sim 0.8$; and the maximum in $\nu F(\nu)$ is $E_{pk-brt} \approx 230$ keV, the FWHM a factor of $\sim 4$ for a nearly symmetric distribution.

For our long-lag sample the mode for the lower power-law is the bin $\alpha = -0.5$ to $-0.2$, with 19 of 25 spectra *flatter* than the median for $\alpha_{brt}$. The difference in modes is $\Delta\alpha_{diff} \sim 0.5$. The upper power-law index peaks in the bin $\beta = -2.9$ to $-2.6$, with 24 of 25 spectra *steeper* than the median for $\beta_{brt}$, and the difference in modes being $\Delta\beta_{diff} \sim 0.5$. The $E_{pk}$ distribution is approximately symmetrical about $E_{pk} \sim 110$ keV – a factor of two lower than $E_{pk-brt}$ – with 24 of 25 $E_{pk}$ values *below* $E_{pk-brt}$. The one outlier at $E_{pk} = 378$ keV is the rising interval for trigger 2193, for which the decaying interval $E_{pk} = 123$ keV. Thus long-lag bursts can be characterized as having, on average, lower $E_{pk}$ and overall spectra which are *more peaked near $E_{pk}$* than are the spectra of bright bursts, due to differences in the distributions of both the low and high energy power-law indices: Long-lag burst spectra are *softer in terms of the high energy power-law* slope as well as lower $E_{pk}$ – effectively, a relative depletion of higher energy photons; but *harder in terms of the low energy power-law slope* – effectively, an enhancement of lower energy photons below $E_{pk}$ relative to bright burst spectra. Thus the pulses in long-lag bursts are distinguished both temporally and spectrally from those in bright bursts.

However, long-lag bursts and dim bursts in general, which have comparable peak fluxes, probably have similar ranges of $E_{pk}$. See Figure 2 and Table 1 of Mallozzi et al. (1995) where burst $E_{pk}$'s decrease to a mean of $\sim 175$ keV for $F_p$ (photons cm$^{-2}$ s$^{-1}$) $\sim 1$–$1.25$. We are not aware of a systematic treatment of the spectra of even dimmer BATSE bursts. Hence to see if there is any difference between long-lag and short-lag dim bursts, we examined the fluence hardness ratios integrated over the whole burst, with the split at $\tau_{lag} = 1$ s in the range $0.5 < F_p < 2.0$. Starting with the superset of 1429 bursts described in section 2.1, these selections resulted in a sample of 93 long-lag and 714 short-lag bursts. Figures 12 and 13 illustrate the distributions of fluence ratios, for channel 3/channel 1 and channel 3/channel 2, respectively. When plotted in the logarithm, all four distributions in the two figures are close to symmetric with modes of –0.2 to –0.1 (note that the area of the short-lag distribution is normalized to that of the long-lag). For both fluence hardness ratios the distribution for the long-lag set (solid histograms) is shifted only slightly to lower values, $\sim 0.25$ in the logarithm, relative to the short-lag set (dotted histograms). Thus via the coarse measurement of hardness ratios we infer that dim long-lag bursts have slightly softer spectra than the remaining bursts in the same peak flux range. Very similar conclusions obtain for other samples when the cuts in peak flux and spectral lag are varied by ~ factor of two.

The spectral shape parameters are plotted by pairs in Figures 14–16. The lower ($\alpha$) and upper ($\beta$) power law indices are plotted in Figure 13; no correlation is apparent. Figures 15 and 16 show $\alpha$ and $\beta$ versus $E_{pk}$, respectively. $E_{pk}$ for our sample appears to be uncorrelated with either power law index.

### 3.3 *Spectral vs. Temporal Parameters*

We return to examining the pulse descriptors for possible correlations. Figures 17–19 illustrate the three spectral shape parameters versus pulse width. Recall that two intervals were fitted per burst, with the division near $\tau_{peak}$. The temporal sequence for symbols in the figures is diamond then square. Figure 17 shows the plot of lower power-law index, $\alpha$, versus width; no clear correlation is suggested. However, we do see that the predominant trend for $\alpha$ within a burst is to increase (10 out of 12 pulses): The portion of the spectrum below the break energy tends to flatten from the rise to the decay portion of a burst. An analogous trend – for the lower power law to flatten as a burst progresses – was reported by Crider et al. (1997). Figure 18 plots the upper power-law index, $\beta$, versus width. The general appearance is that $\beta$ values are nearly flat (reflecting the narrow distribution in Figure 10), and independent of width. $\beta$ does not appear to exhibit a clear direction of temporal evolution within a burst. Similarly, in Figure 19 $E_{pk}$ appears to be uncorrelated with pulse width. Note that the lack of correlation of any of the three spectral parameters with width is over a dynamic range of almost a factor of ten in the latter parameter.

Figures 20–22 illustrate a similar picture for the three spectral shape parameters versus pulse asymmetry, except that there is a suggestion that a flatter $\alpha$ slope is correlated with a more asymmetric pulse. Figure 20a plots $\alpha$ for both spectral intervals. In Figure 20b the $\alpha$ values are averaged and a straight line fitted to the points, $\alpha = 2.58\kappa - 1.62$. The Spearman rank coefficient is 0.70 (P ~ $7 \times 10^{-3}$), a mildly suggestive trend given the scatter evident in the plot. Finally, $\beta$ (Figure 21) and $E_{pk}$ (Figure 22) appear uncorrelated with asymmetry.

The overall conclusion is that the five fundamental temporal/spectral parameters that we consider, which adequately describe observed pulses in long-lag, wide-pulse GRBs – $w$, $\kappa$, $\alpha$, $\beta$ and $E_{pk}$ – do not show compelling evidence for any pair-wise correlation, except for a possible weak correlation between $\alpha$ and $\kappa$. This implies that at least four to five independent physical parameters are required to determine pulse behavior in the energy band ~ 25–1000 keV.

## 4. DISCUSSION

We first review the evidence that long-lag bursts form either a separate subclass of GRBs or perhaps one end of an important part of the GRB physical phase space. Then we discuss the observational and theoretical bases for conducting our search for the number of independent parameters needed to describe their pulses, the primary results from this analysis, and additional evidence from other sources. We quantify the importance of observations in the Swift era, in terms of expected GRB yield and afterglow follow-up observations. We conclude with a short description of simulations that may narrow the physical parameter space that determines luminosity and energy release for these bursts, and how combinations of physical parameters may be related to the observables we have studied.

The work of Stern, Poutanen, & Svensson (1999) first suggested a group of "simple" bursts with peak fluxes near the BATSE trigger threshold: their complexity parameter indicated that the average profile of dim bursts were less complex than that of bright bursts. In fact, amongst the low peak flux bursts are those with long spectral lag and few, wide pulses – the group we studied in this work. Their spectral lags range from a few tenths of a second to several seconds with pulse widths ranging from ten to tens of seconds (Figures 5–8). Their spectra are slightly softer than other bursts within the same peak flux range (Figures 9–12), with the mean for $E_{pk}$ a factor of two lower than for bright bursts.

The most obvious feature of the long-lag burst population is that their proportion within long-duration bursts ($T_{90} > 2$ s) increases from negligible among bright BATSE bursts to ~ 50% at trigger threshold (Figure 3, Norris 2002). Taken together with the fact that the first such burst with a determined redshift was GRB 980425/SN 1998bw with a source distance of ~ 38 Mpc (Galama et al. 1998), this argues that long-lag bursts are probably relatively nearby. Nearby burst sources were predicted by several groups following the discovery of SN 1998bw, with their low luminosities being partly attributed to low Lorentz factors ($\Gamma \sim 2$–5: Kulkarni et al. 1998; Woosley & MacFadyen 1999; Salmonson 2001), and partly to large off-axis viewing angle (Nakamura 1999; Salmonson 2001). The fact that the long-lag bursts have peak fluxes two orders of magnitude lower than do the brightest bursts argues that they are intrinsically underluminous. Perhaps the least contrived alternative – that long-lag bursts are "ordinary" bursts at redshifts of ~ 10 – does not fit any of the defining attributes: (1) their spectra are similar to those of other dim bursts, with $E_{pk} \sim 100$–200 keV, not ~ 10–20 keV; (2) their time profiles are not stretched versions of bright bursts – only the (very few) individual pulses are long; and (3) general relativistic effects for $z \sim 10$ would result in luminosity diminished by factor of a few, not ~ 100 (see Lamb & Reichart 2000). Thus our working hypothesis has been that long-lag bursts are underluminous, and the question is how many factors determine their low luminosity and low total energy output.

The question of how long-lag bursts might fit into a more general classification scheme has been studied by a variety of authors who identified three possible GRB classes using statistical clustering tests (Mukherjee et al. 1998; Horvath 1998; Balastegui, Ruiz-Lapuente, & Canal 2001; Rajaniemi & Mahonen 2002; Hakkila et al. 2003). In addition to the well-known long and short duration classes (Kouveliotou et al. 1993), these studies generally identify a separate class characterized by soft spectra, intermediate durations and intermediate fluences.

Hakkila et al. (2003) hypothesized that the intermediate class was due to a bias favoring detection of bursts varying on short trigger timescales. They noted that bursts with large peak fluxes but small fluences were included in the BATSE sample whereas bursts with small peak fluxes but large fluences were not. This truncation is caused by the short (~ 1 s) timescale of the BATSE trigger. In order to eliminate this bias, these authors introduced a dual timescale peak flux to avoid favoring either rapidly varying or slowly varying bursts with respect to one another. The dual timescale trigger is a combination of a short trigger (~ 1 s) with a very long one (>> 1 s). The authors applied this criterion to the BATSE catalog and truncated it below a dual timescale threshold at which the sample was considered to be complete. Most of the bursts in the intermediate class were removed as a result of introducing this threshold, and data mining tools were unable to identify a well-defined intermediate class from the remaining data. In drawing their conclusion that the trigger timescale caused the inclusion of an inordinately large number of intermediate bursts, Hakkila et al. had an underlying assumption that bursts with small numbers of pulses (short and intermediate bursts have this characteristic) concentrate a large amount of flux near their pulse peaks; this is what they believed caused these bursts to preferentially trigger on short trigger timescales. Since the properties of intermediate bursts are similar to those of faint long bursts (e.g. in fluence, duration, peak flux, and spectral hardness), the authors also started from the assumption that long and intermediate bursts might be intrinsically similar.

The existence of slowly varying, long-lag bursts, combined with the characteristics that these bursts share with the intermediate class, suggests that long lags might characterize many bursts in the intermediate class. It is also an indication that there might be measurable intrinsic differences between the long and intermediate burst classes that are best characterized by attributes other than fluence, duration, and spectral hardness. These attributes might include spectral lags, variability, and the internal luminosity function (Hakkila et al. 2004). Automated classification is needed to determine whether or not these additional attributes can be used to delineate the intermediate class from the long class.

We note that the spectral evolution of a pulse is implicit in the energy dependences of pulse width ($w \propto E^{-0.4}$, eq. [7]) and shape, which we have modeled as the asymmetry parameter ($\kappa$, eqs. [3] and [4]). For our sample of wide-pulse bursts a slight tendency is evident in Figure 2 for the distribution in κ to narrow at higher energies, with the average κ ~ 0.4 independent of energy. Moreover, spectral lag is proportional to pulse width ($\tau_{lag13} \propto w^{1.42}$, eq. [8]). Clearly, the spectral evolution of wide pulses is faster than that of narrow pulses. The details of pulse evolution are determined by a combination of properties intrinsic to the emitting region, including energization and cooling processes and jet profile, and extrinsic properties, such as viewing angle and related absorption. Hence, a pimary motivation of this work was to characterize the diversity of long-lag pulse shapes, which must be explicable in terms of a combination of such intrinsic and extrinsic properties. Proposed explanations for the fact that pulse emission peaks and decays more quickly at higher energies, giving rise to spectral lag, include both cooling and viewing angle effects (Fenimore, Madras, Nayakshin 1996; Sari & Piran 1997; Soderberg & Fenimore 2001).

Since the pulses in long-lag bursts are very long – 100 times longer than pulses in bright, complex bursts – sufficient pulse definition is available to make for an easy study. In addition, these simple bursts – with presumed lower total energy, slower evolution, lower bulk Lorentz factor – may have relatively simpler physics. Thus they may be the ideal type of burst to model first in order to gain understanding of jet dynamics. Observationally, we expect that the minimum number of free parameters required to model bursts should be at least the current number of luminosity/total energy indicators that have been proposed (e.g., spectral lag [Norris, Marani & Bonnell 2000], variability [Reichart et al. 2001], $E_{pk}$ [Amati et al. 2002]), if in fact such parameters are completely independent. It is possible that a single physical variable may drive two observed quantities that only appear to be independent, due to irregular dependence on the physical variable, high sensitivity to other "hidden" variables, and observational noise.

These considerations are timely since most theoretical problems concerning primal energy source, GRB jet structure (opening angle, profiles of Lorentz factor and matter/field density), distance of emission region from central source, viewing angle, and even matter versus field flow are unresolved (e.g., Lyutikov & Blandford 2004). Such a large phase space is commensurate with our results, discussed and summarized in the previous section. Figures 14–16 show that the three spectral shape parameters – α, β, and $E_{pk}$ – appear mutually uncorrelated, and Figures 17–22 show little evidence for any correlation between the temporal shape parameters – $w$ and κ (which also appear to be uncorrelated, Figure 2) – and the spectral shape parameters. This suggests that ≳ 5 independent parameters are required to determine pulse temporal and spectral behavior in long-lag bursts. However, as anticipated, pulse width is strongly correlated with spectral lag at some energies (Figure 8a); hence these two parameters may be viewed as mutual surrogates in formulations for estimating GRB luminosity and total energy. Thus the old adage describing the diversity of GRBs in general, "If you've seen one

GRB, you've seen one GRB" (attributed to Bill Paciesas), seems to apply as well to the long-lag, wide-pulse subclass. Even though these bursts tend to have just one to a few pulses, pulse temporal and spectral shape varies considerably from burst to burst.

Additional observations provide evidence that more than one parameter determines luminosity, even within a burst. Our own (unpublished) studies of bright, complex bursts indicate that within a given burst individual pulses tend to have comparable widths and spectral lags, but that these quantities are not closely correlated with pulse intensity. A similar result has been reported recently by Chen et al. (2004) from analysis of BATSE time-to-spill data. Moreover, infrequently (two cases identified so far) delayed long-lag emission is observed in bursts with otherwise short-lag behavior. This may be explained as external shocks following the main prompt emission from internal shocks (Hakkila & Giblin 2004; Hakkila et al. 2004). In the case of subluminous GRB 031203, whose source was determined to be at a low redshift ($z$ = 0.106; Prochaska et al. 2004), the spectral lag measure is relatively long ($\tau_{lag} \sim 0.24 \pm 0.12$), placing this burst like GRB 980425 in an outlier position in the peak-luminosity – spectral-lag diagram. The implications from hard X-ray and radio studies are that this burst manifested a little collimation, rather than a narrow jet-like, emission pattern, and that a large population of nearby low luminosity bursts exists with low or negligible optical afterglow emission (Soderberg, A. M., et al. 2004; Sazonov, Lutovinov, & Sunyaev 2004). Thus, while there is evidence that long-lag, wide-pulse bursts are subluminous and perhaps have more nearly spherical emission, their nature and site(s) of this emission with respect to the central source, and its relation to short-lag emission are unclear, with several factors unresolved.

Swift should see a sizeable yield of dim, long-lag bursts within a factor of ten of its detection threshold. For BATSE the proportion of long-lag bursts at threshold was ~ 50%. Swift's effective area (at half maximum) is broadly peaked (~ 15–150 keV), a factor of two below BATSE (~ 30–350 keV). Since these bursts are relatively soft, with $E_{pk}$ clustering in the range 100–200 keV and upper power-law indices steeper by $\Delta\beta_{diff} \sim 0.5$ compared to spectra of bright bursts, Swift will be slightly more sensitive to their emission than was BATSE (cf. Figure 2 and Figure 7 of Band 2003). If even softer long-lag bursts were undetected just below BATSE threshold, then Swift will have a much larger yield than did BATSE. Furthermore, BATSE's longest trigger timescale was 1.024 s, whereas the Swift BAT trigger includes much longer accumulation times, up to 16 s (McLean et al. 2004), more commensurate with the long pulses in our long-lag burst sample. Even if optical afterglows are not prevalent for these bursts, real-time alerts and accurate localizations provided by Swift's X-Ray Telescope will enable detection and study of supernovae which should be expected to be associated with long-lag GRBs, if in fact these bursts are underluminous and relatively nearby.

In future work we will address some aspects of the multi-parameter problem using simulations of jet kinematics aimed at modeling the long-lag, wide-pulse bursts, where the free parameters include: angular extent of jet; profile of mass density; shell thickness; profile of Lorentz factor; source optically thin distance, and observer angle with respect to jet axis. The intent is to constrain how the physical parameters govern the variation of luminosity with time and energy.

TABLE 1. Long-lag Burst Sample

| Trigger # | $F_p{}^a$ | $T_{90}{}^b$ | MER Comment $^c$ | Structure Comments |
|---|---|---|---|---|
| 764 $^d$ | 1.014 | 33.79 | Complete | 1 peaked pulse, no ch 4 |
| 1039 $^d$ | 1.377 | 16.89 | Complete $^f$ | 1 peaked pulse, no ch 4 |
| 1406 | 2.129 | 19.96 | Complete $^f$ | 1 smooth pulse, low ch 4 |
| 2193 | 1.451 | 114.68 | Complete $^f$ | 2 smooth pulses, blended, low post emission, med ch 4 |
| 2197 | 0.957 | 30.72 | no data | 1 smooth pulse, very low ch 4 |
| 2387 | 3.584 | 33.28 | Complete $^f$ | 1 smooth pulse, low ch 4 |
| 2665 | 1.909 | 18.94 | gaps | 1 peaked pulse, no ch 4 |
| 2711 $^d$ | 0.875 | 122.36 | no data | 3 pulses, very low ch 4 |
| 2863 $^d$ | 1.719 | 37.88 | Complete $^f$ | 2 smooth pulses, blended, low ch 4 |
| 3256 | 1.715 | 75.26 | Complete $^f$ | 2 smooth pulses, 2$^{nd}$ low amp, very low ch 4 |
| 3257 | 2.624 | 40.44 | gap, ch 11 $^f$ | 1 smooth pulse, very low ch 4 |
| 5387 | 1.250 | 40.44 | gaps, ch 1 & 11 | 1 smooth, 2$^{nd}$ low amp, very low ch 4 |
| 5415 | 1.302 | 27.13 | no data | 1 smooth pulse, very low ch 4 |
| 6147 $^d$ | 1.078 | 40.96 | Complete $^f$ | 2 smooth pulses, low 2$^{nd}$ pulse blended, very low ch 4 |
| 6414 $^d$ | 1.930 | 37.88 | Complete $^f$ | 1 peaked pulse, very low ch 4 |
| 6504 | 2.253 | 25.08 | Complete $^f$ | 1 smooth pulse, low post emission, very low ch 4 |
| 6526 | 1.500 | 900.00 $^e$ | Complete | 1 smooth pulse, very low ch 4; longest single pulse |
| 6598 | 1.167 | 13.82 | no data | 1 smooth pulse, no ch 4 |
| 6625 $^d$ | 1.679 | 25.08 | gap, ch 9 $^f$ | 1 peaked pulse, precursors, no ch 4 |
| 6630 $^d$ | 15.336 | 19.45 | Complete | 4 or more pulses, blended, high ch 4; brightest in sample |
| 6707 | 0.911 | 30.72 | no data | 1 smooth pulse, no ch 4 |
| 7087 | 1.202 | 43.00 | no data | 1 smooth pulse, very low ch 4 |
| 7156 | 1.467 | 18.94 | no data | 1 smooth pulse, very low ch 4 |
| 7293 | 2.725 | 32.76 | gap, ch 11 $^f$ | 1 smooth pulse, low ch 4 |
| 7403 | 1.221 | 40.96 | no data | 1 smooth pulse, very low ch 4 |
| 7588 | 2.063 | 18.43 | Complete $^f$ | 1 smooth pulse, very low ch 4 |
| 7648 $^d$ | 1.530 | 19.96 | gap, ch 12 $^f$ | 1 smooth pulse, very low ch 4 |
| 7969 $^d$ | 3.077 | 16.89 | Complete $^f$ | 3 smooth pulses, 1$^{st}$ two blended, very low ch 4 |
| 8049 | 1.573 | 70.14 | Complete $^f$ | 3 smooth pulses, little blend, very low ch 4 |

$^a$ Peak Flux, photons cm$^{-2}$ s$^{-1}$ (50–300 keV)
$^b$ Interval (s) between times when 5% and 95% of counts accumulate (>25 keV)
$^c$ MER = Medium Energy Resolution, 16 channel data; comments refer to burst part of MER data record
$^d$ eliminated after pulse fitting due to discovered overlapping pulses
$^e$ visual estimate
$^f$ included in MER subsample for spectral analysis

TABLE 2. Temporal Fit Parameters

| Trig# | Ch | $\chi^2/\nu$ | A | $t_s$ | $t_{eff}$ | $\tau_{peak}$ | $\tau_1$ | $\tau_2$ | w | $\kappa$ |
|---|---|---|---|---|---|---|---|---|---|---|
| 764 | 1 | 0.929 | | | | | | | | |
| | | | 582.20 | -4.79 | -3.68 | 5.81 | 7.23 | 6.62 | 15.07 | 0.44 |
| | | | 21.81 | 0.76 | 0.76 | 1.42 | 2.84 | 0.79 | 2.18 | 0.04 |
| | | | 98.16 | 24.00 | 24.00 | 0.27 | 0.00 | 48.65 | 49.20 | 0.99 |
| | | | 22.06 | 0.06 | 0.06 | 4.37 | 0.05 | 15.63 | 18.02 | 0.17 |
| | 2 | 1.173 | | | | | | | | |
| | | | 1023.37 | -8.96 | -5.43 | 6.77 | 37.66 | 2.82 | 11.14 | 0.25 |
| | | | 27.06 | 1.12 | 1.12 | 1.94 | 13.35 | 0.36 | 1.73 | 0.02 |
| | | | 186.78 | 10.30 | 10.41 | 2.40 | 0.55 | 11.45 | 15.69 | 0.73 |
| | | | 29.55 | 0.63 | 0.63 | 2.15 | 0.93 | 3.14 | 5.33 | 0.15 |
| | 3 | 1.392 | | | | | | | | |
| | | | 738.94 | -10.51 | -7.55 | 7.50 | 26.55 | 4.12 | 13.77 | 0.30 |
| | | | 21.25 | 1.10 | 1.10 | 1.86 | 8.97 | 0.45 | 1.86 | 0.02 |
| 1406 | 1 | 1.106 | | | | | | | | |
| | | | 1014.49 | -2.98 | -1.48 | 5.86 | 10.87 | 4.97 | 13.08 | 0.38 |
| | | | 3.62 | 0.02 | 0.02 | 0.03 | 0.06 | 0.02 | 0.06 | 0.00 |
| | | | 110.85 | 10.71 | 10.87 | 6.34 | 0.73 | 57.71 | 69.50 | 0.83 |
| | | | 1.42 | 0.14 | 0.14 | 0.41 | 0.09 | 1.32 | 1.72 | 0.01 |
| | 2 | 1.112 | | | | | | | | |
| | | | 1378.02 | -2.00 | -1.07 | 4.54 | 6.16 | 4.84 | 11.37 | 0.43 |
| | | | 18.38 | 0.20 | 0.20 | 0.43 | 0.90 | 0.29 | 0.77 | 0.01 |
| | | | 107.63 | 9.60 | 9.80 | 5.99 | 1.00 | 38.23 | 49.08 | 0.78 |
| | | | 16.01 | 0.96 | 0.96 | 3.34 | 1.06 | 8.14 | 11.67 | 0.08 |
| | 3 | 1.249 | | | | | | | | |
| | | | 1151.90 | -1.48 | -1.07 | 3.24 | 2.38 | 5.60 | 10.63 | 0.53 |
| | | | 15.17 | 0.10 | 0.10 | 0.21 | 0.27 | 0.16 | 0.38 | 0.01 |
| | 4 | 1.012 | | | | | | | | |
| | | | 198.93 | -1.32 | -1.20 | 1.54 | 0.66 | 4.16 | 6.70 | 0.62 |
| | | | 11.56 | 0.21 | 0.21 | 0.53 | 0.41 | 0.54 | 1.09 | 0.06 |
| 2193 | 1 | 1.054 | | | | | | | | |
| | | | 984.94 | -4.96 | -3.27 | 19.60 | 9.18 | 49.39 | 81.51 | 0.61 |
| | | | 29.37 | 1.19 | 1.19 | 2.98 | 2.45 | 4.24 | 7.88 | 0.03 |
| | | | 207.66 | 98.03 | 99.56 | 16.87 | 8.38 | 40.40 | 67.86 | 0.60 |
| | | | 29.21 | 5.11 | 5.11 | 11.91 | 10.37 | 15.43 | 29.55 | 0.12 |
| | 2 | 1.069 | | | | | | | | |
| | | | 2769.87 | -6.91 | -3.83 | 19.37 | 19.06 | 26.44 | 55.44 | 0.48 |
| | | | 31.48 | 0.66 | 0.66 | 1.33 | 2.16 | 0.93 | 2.33 | 0.01 |

|   |   |   | 318.91 | 68.87 | 83.20 | 34.81 | 131.43 | 18.37 | 62.83 | 0.29 |
|---|---|---|---|---|---|---|---|---|---|---|
|   |   |   | 25.23 | 15.27 | 15.27 | 25.93 | 131.27 | 6.26 | 26.23 | 0.07 |
|   | 3 | 1.149 |   |   |   |   |   |   |   |   |
|   |   |   | 5671.40 | -5.56 | -3.12 | 12.78 | 15.94 | 14.53 | 33.11 | 0.44 |
|   |   |   | 40.08 | 0.27 | 0.27 | 0.52 | 1.05 | 0.28 | 0.78 | 0.01 |
|   |   |   | 193.61 | 21.62 | 72.67 | 34.67 | 1437.11 | 5.11 | 42.18 | 0.12 |
|   |   |   | 25.56 | 35.83 | 35.83 | 55.17 | 1716.08 | 2.46 | 24.30 | 0.04 |
|   | 4 | 0.936 |   |   |   |   |   |   |   |   |
|   |   |   | 967.89 | -4.75 | -3.38 | 7.12 | 8.97 | 8.03 | 18.37 | 0.44 |
|   |   |   | 33.44 | 0.83 | 0.83 | 1.58 | 3.20 | 0.84 | 2.36 | 0.03 |
| 2197 | 1 | 1.019 |   |   |   |   |   |   |   |   |
|   |   |   | 347.64 | -7.51 | -2.98 | 12.43 | 38.87 | 7.40 | 23.60 | 0.31 |
|   |   |   | 10.18 | 1.77 | 1.77 | 3.08 | 13.36 | 0.86 | 3.36 | 0.03 |
|   | 2 | 0.900 |   |   |   |   |   |   |   |   |
|   |   |   | 577.71 | -3.34 | -1.80 | 6.98 | 10.56 | 6.87 | 16.77 | 0.41 |
|   |   |   | 11.77 | 0.51 | 0.51 | 0.94 | 2.23 | 0.45 | 1.34 | 0.02 |
|   | 3 | 1.109 |   |   |   |   |   |   |   |   |
|   |   |   | 657.05 | -1.64 | -1.13 | 3.69 | 3.07 | 5.76 | 11.41 | 0.51 |
|   |   |   | 13.15 | 0.20 | 0.20 | 0.41 | 0.57 | 0.30 | 0.72 | 0.02 |
|   | 4 | 0.751 |   |   |   |   |   |   |   |   |
|   |   |   | 81.97 | -1.03 | -1.02 | 0.14 | 0.01 | 2.95 | 3.23 | 0.92 |
|   |   |   | 28.08 | 0.03 | 0.03 | 0.53 | 0.05 | 1.00 | 1.47 | 0.28 |
| 2387 | 1 | 1.175 |   |   |   |   |   |   |   |   |
|   |   |   | 1333.14 | -1.21 | -0.13 | 7.74 | 6.47 | 12.03 | 23.86 | 0.50 |
|   |   |   | 12.29 | 0.19 | 0.19 | 0.39 | 0.55 | 0.30 | 0.71 | 0.01 |
|   | 2 | 1.198 |   |   |   |   |   |   |   |   |
|   |   |   | 2032.13 | -2.35 | -0.83 | 7.68 | 9.98 | 8.48 | 19.59 | 0.43 |
|   |   |   | 13.77 | 0.15 | 0.15 | 0.29 | 0.61 | 0.15 | 0.43 | 0.01 |
|   | 3 | 1.051 |   |   |   |   |   |   |   |   |
|   |   |   | 2315.19 | -3.24 | -1.40 | 6.90 | 13.54 | 5.64 | 15.13 | 0.37 |
|   |   |   | 15.70 | 0.16 | 0.16 | 0.29 | 0.85 | 0.11 | 0.36 | 0.01 |
|   | 4 | 0.944 |   |   |   |   |   |   |   |   |
|   |   |   | 200.49 | -3.89 | -2.48 | 5.51 | 10.24 | 4.67 | 12.29 | 0.38 |
|   |   |   | 9.52 | 1.06 | 1.06 | 1.87 | 5.31 | 0.74 | 2.41 | 0.04 |
| 2665 | 1 | 0.931 |   |   |   |   |   |   |   |   |
|   |   |   | 588.98 | -0.67 | -0.51 | 2.77 | 0.86 | 9.96 | 14.70 | 0.68 |
|   |   |   | 15.66 | 0.16 | 0.16 | 0.43 | 0.25 | 0.63 | 1.10 | 0.03 |
|   | 2 | 1.256 |   |   |   |   |   |   |   |   |
|   |   |   | 1044.98 | -0.95 | -0.64 | 2.19 | 1.89 | 3.33 | 6.67 | 0.50 |
|   |   |   | 21.98 | 0.15 | 0.15 | 0.28 | 0.41 | 0.18 | 0.46 | 0.02 |

|  |  |  |  |  |  |  |  |  |  |
|---|---|---|---|---|---|---|---|---|---|
|  | 3 | 1.003 |  |  |  |  |  |  |  |
|  |  |  | 908.41 | -1.99 | -1.18 | 2.08 | 7.32 | 1.15 | 3.82 | 0.30 |
|  |  |  | 26.55 | 0.30 | 0.30 | 0.51 | 2.47 | 0.12 | 0.51 | 0.02 |
| 2711 | 1 | 1.308 |  |  |  |  |  |  |  |
|  |  |  | 414.50 | -12.10 | -2.97 | 6.73 | 233.32 | 1.08 | 8.34 | 0.13 |
|  |  |  | 18.25 | 2.18 | 2.18 | 3.64 | 98.32 | 0.20 | 1.78 | 0.01 |
|  |  |  | 124.91 | 4.84 | 7.49 | 11.83 | 18.32 | 11.45 | 28.19 | 0.41 |
|  |  |  | 9.69 | 4.59 | 4.59 | 7.44 | 18.07 | 3.31 | 10.15 | 0.09 |
|  |  |  | 905.09 | 17.45 | 82.86 | 26.14 | 3693.18 | 2.27 | 28.92 | 0.08 |
|  |  |  | 10.41 | 2.16 | 2.16 | 3.37 | 249.77 | 0.07 | 1.00 | 0.00 |
|  | 2 | 1.247 |  |  |  |  |  |  |  |
|  |  |  | 562.72 | -2.62 | -1.68 | 3.60 | 6.80 | 3.02 | 7.99 | 0.38 |
|  |  |  | 17.11 | 0.45 | 0.45 | 0.82 | 2.35 | 0.34 | 1.08 | 0.03 |
|  |  |  | 144.60 | 10.08 | 11.07 | 8.10 | 5.79 | 14.31 | 26.94 | 0.53 |
|  |  |  | 8.85 | 1.42 | 1.42 | 2.88 | 3.52 | 2.52 | 5.64 | 0.06 |
|  |  |  | 460.98 | -51.89 | 79.92 | 28.41 | 19299. | 1.33 | 29.23 | 0.05 |
|  |  |  | 8.47 | 3.87 | 3.87 | 5.88 | 1239. | 0.05 | 1.17 | 0.00 |
|  | 3 | 1.018 |  |  |  |  |  |  |  |
|  |  |  | 458.90 | -2.15 | -1.66 | 2.41 | 3.26 | 2.59 | 6.07 | 0.43 |
|  |  |  | 17.25 | 0.31 | 0.31 | 0.58 | 1.24 | 0.29 | 0.84 | 0.03 |
|  |  |  | 105.09 | 13.57 | 13.57 | 0.02 | 0.00 | 14.36 | 14.39 | 1.00 |
|  |  |  | 13.05 | 0.00 | 0.00 | 0.22 | 0.00 | 3.01 | 3.05 | 0.03 |
|  |  |  | 73.83 | 79.94 | 86.03 | 17.22 | 51.31 | 10.59 | 33.15 | 0.32 |
|  |  |  | 6.76 | 7.73 | 7.73 | 13.43 | 56.17 | 3.80 | 14.67 | 0.08 |
| 3256 | 1 | 1.156 |  |  |  |  |  |  |  |
|  |  |  | 390.48 | -1.81 | -1.13 | 4.95 | 4.08 | 7.77 | 15.35 | 0.51 |
|  |  |  | 14.05 | 0.44 | 0.44 | 0.88 | 1.22 | 0.72 | 1.67 | 0.03 |
|  |  |  | 168.25 | -0.14 | 28.73 | 33.91 | 455.65 | 8.65 | 47.41 | 0.18 |
|  |  |  | 8.92 | 4.46 | 4.46 | 6.98 | 89.29 | 0.91 | 5.59 | 0.01 |
|  | 2 | 1.077 |  |  |  |  |  |  |  |
|  |  |  | 790.08 | -1.27 | -0.83 | 3.00 | 2.66 | 4.46 | 9.02 | 0.49 |
|  |  |  | 17.12 | 0.16 | 0.16 | 0.32 | 0.48 | 0.23 | 0.57 | 0.02 |
|  |  |  | 123.19 | -12.76 | 30.15 | 34.07 | 1009.24 | 5.87 | 42.93 | 0.14 |
|  |  |  | 7.34 | 5.42 | 5.42 | 8.21 | 184.30 | 0.65 | 5.24 | 0.01 |
|  | 3 | 1.098 |  |  |  |  |  |  |  |
|  |  |  | 1068.55 | -0.79 | -0.60 | 1.52 | 1.11 | 2.64 | 5.01 | 0.53 |
|  |  |  | 23.75 | 0.10 | 0.10 | 0.21 | 0.27 | 0.14 | 0.35 | 0.02 |
|  | 4 | 0.830 |  |  |  |  |  |  |  |

|      |   |       |         |       |       |      |        |       |       |      |
|------|---|-------|---------|-------|-------|------|--------|-------|-------|------|
|      |   |       | 98.40   | -5.04 | -1.52 | 2.21 | 108.49 | 0.30  | 2.65  | 0.11 |
|      |   |       | 17.35   | 1.18  | 1.18  | 1.77 | 57.48  | 0.10  | 0.93  | 0.02 |
| 3257 | 1 | 1.112 |         |       |       |      |        |       |       |      |
|      |   |       | 484.25  | -0.44 | -0.21 | 5.22 | 1.15   | 25.84 | 35.08 | 0.74 |
|      |   |       | 9.46    | 0.17  | 0.17  | 0.57 | 0.23   | 1.20  | 1.83  | 0.02 |
|      | 2 | 1.284 |         |       |       |      |        |       |       |      |
|      |   |       | 1076.31 | -0.38 | -0.18 | 3.93 | 1.01   | 16.84 | 23.70 | 0.71 |
|      |   |       | 11.97   | 0.07  | 0.07  | 0.22 | 0.11   | 0.38  | 0.62  | 0.01 |
|      | 3 | 1.813 |         |       |       |      |        |       |       |      |
|      |   |       | 1711.86 | -0.47 | -0.28 | 2.77 | 0.96   | 9.04  | 13.73 | 0.66 |
|      |   |       | 17.17   | 0.05  | 0.05  | 0.14 | 0.09   | 0.17  | 0.32  | 0.01 |
|      | 4 | 0.804 |         |       |       |      |        |       |       |      |
|      |   |       | 172.48  | -1.28 | -1.12 | 1.71 | 0.86   | 4.06  | 6.85  | 0.59 |
|      |   |       | 13.61   | 0.26  | 0.26  | 0.64 | 0.57   | 0.72  | 1.43  | 0.07 |
| 5387 | 1 | 1.080 |         |       |       |      |        |       |       |      |
|      |   |       | 737.02  | -3.40 | -2.16 | 7.96 | 7.64   | 11.09 | 23.05 | 0.48 |
|      |   |       | 6.82    | 0.08  | 0.08  | 0.12 | 0.17   | 0.13  | 0.30  | 0.00 |
|      |   |       | 156.15  | 34.15 | 34.28 | 3.71 | 0.66   | 22.46 | 29.16 | 0.77 |
|      |   |       | 6.09    | 0.39  | 0.39  | 0.57 | 0.19   | 1.34  | 1.94  | 0.02 |
|      | 2 | 1.179 |         |       |       |      |        |       |       |      |
|      |   |       | 1395.52 | -3.18 | -1.86 | 6.59 | 8.76   | 7.15  | 16.65 | 0.43 |
|      |   |       | 29.25   | 0.48  | 0.48  | 0.89 | 1.89   | 0.46  | 1.32  | 0.02 |
|      |   |       | 149.94  | 33.54 | 34.45 | 5.91 | 5.57   | 8.34  | 17.24 | 0.48 |
|      |   |       | 26.10   | 3.11  | 3.11  | 5.94 | 9.23   | 4.50  | 10.93 | 0.16 |
|      | 3 | 0.925 |         |       |       |      |        |       |       |      |
|      |   |       | 2194.55 | -1.79 | -1.33 | 3.19 | 2.76   | 4.83  | 9.69  | 0.50 |
|      |   |       | 38.38   | 0.19  | 0.19  | 0.36 | 0.53   | 0.23  | 0.58  | 0.02 |
|      | 4 | 0.984 |         |       |       |      |        |       |       |      |
|      |   |       | 201.69  | -2.36 | -1.89 | 1.97 | 3.27   | 1.81  | 4.57  | 0.40 |
|      |   |       | 33.02   | 1.02  | 1.02  | 2.06 | 5.25   | 0.99  | 2.98  | 0.14 |
| 5415 | 1 | 0.997 |         |       |       |      |        |       |       |      |
|      |   |       | 476.31  | -3.70 | -1.83 | 7.26 | 13.67  | 6.11  | 16.14 | 0.38 |
|      |   |       | 13.36   | 0.81  | 0.81  | 1.47 | 4.18   | 0.60  | 1.94  | 0.03 |
|      | 2 | 1.153 |         |       |       |      |        |       |       |      |
|      |   |       | 755.65  | -2.56 | -1.33 | 4.96 | 8.75   | 4.37  | 11.27 | 0.39 |
|      |   |       | 14.96   | 0.37  | 0.37  | 0.67 | 1.81   | 0.28  | 0.89  | 0.02 |
|      | 3 | 0.791 |         |       |       |      |        |       |       |      |
|      |   |       | 740.46  | -3.08 | -1.94 | 4.17 | 8.54   | 3.31  | 9.01  | 0.37 |
|      |   |       | 15.54   | 0.35  | 0.35  | 0.62 | 1.89   | 0.23  | 0.76  | 0.02 |
| 6147 | 1 | 0.787 |         |       |       |      |        |       |       |      |

|  |  |  | 412.41 | -4.49 | -2.44 | 7.91 | 14.95 | 6.63 | 17.56 | 0.38 |
|---|---|---|---|---|---|---|---|---|---|---|
|  |  |  | 13.44 | 1.05 | 1.05 | 2.01 | 5.66 | 0.91 | 2.84 | 0.03 |
|  |  |  | 131.72 | 14.54 | 19.12 | 12.26 | 39.87 | 7.12 | 23.03 | 0.31 |
|  |  |  | 11.79 | 5.27 | 5.27 | 8.82 | 39.59 | 2.37 | 9.41 | 0.07 |
|  | 2 | 0.918 |  |  |  |  |  |  |  |  |
|  |  |  | 744.76 | -4.10 | -2.13 | 6.43 | 15.45 | 4.56 | 13.19 | 0.35 |
|  |  |  | 13.95 | 0.34 | 0.34 | 0.59 | 2.05 | 0.22 | 0.76 | 0.01 |
|  |  |  | 116.08 | 22.19 | 22.65 | 4.16 | 2.54 | 8.37 | 14.99 | 0.56 |
|  |  |  | 11.89 | 0.96 | 0.96 | 1.97 | 2.07 | 2.24 | 4.57 | 0.08 |
|  | 3 | 0.801 |  |  |  |  |  |  |  |  |
|  |  |  | 561.03 | -2.26 | -1.55 | 3.65 | 4.59 | 4.13 | 9.44 | 0.44 |
|  |  |  | 15.45 | 0.33 | 0.33 | 0.63 | 1.27 | 0.34 | 0.95 | 0.03 |
| 6504 | 1 | 1.079 |  |  |  |  |  |  |  |  |
|  |  |  | 528.74 | -1.52 | -0.91 | 6.49 | 3.37 | 14.96 | 25.47 | 0.59 |
|  |  |  | 2.74 | 0.04 | 0.04 | 0.07 | 0.06 | 0.11 | 0.20 | 0.00 |
|  |  |  | 74.37 | 30.35 | 30.48 | 4.29 | 0.63 | 31.04 | 38.89 | 0.80 |
|  |  |  | 2.29 | 0.21 | 0.21 | 0.56 | 0.15 | 2.22 | 2.92 | 0.02 |
|  | 2 | 0.957 |  |  |  |  |  |  |  |  |
|  |  |  | 1100.47 | -1.49 | -0.75 | 4.96 | 4.54 | 7.18 | 14.67 | 0.49 |
|  |  |  | 16.24 | 0.19 | 0.19 | 0.40 | 0.60 | 0.31 | 0.74 | 0.01 |
|  |  |  | 87.59 | 30.13 | 30.44 | 5.44 | 1.61 | 20.57 | 29.94 | 0.69 |
|  |  |  | 11.50 | 1.38 | 1.38 | 3.90 | 2.10 | 7.64 | 12.34 | 0.12 |
|  | 3 | 1.006 |  |  |  |  |  |  |  |  |
|  |  |  | 1763.20 | -1.31 | -0.69 | 3.25 | 3.98 | 3.75 | 8.49 | 0.44 |
|  |  |  | 22.18 | 0.13 | 0.13 | 0.24 | 0.47 | 0.13 | 0.36 | 0.01 |
|  | 4 | 0.979 |  |  |  |  |  |  |  |  |
|  |  |  | 138.78 | -4.77 | -2.36 | 3.77 | 29.79 | 1.28 | 5.78 | 0.22 |
|  |  |  | 18.14 | 1.56 | 1.56 | 2.45 | 21.60 | 0.41 | 2.13 | 0.04 |
| 6526 | 1 | 4.064 |  |  |  |  |  |  |  |  |
|  |  |  | 7886.56 | -7.29 | -2.75 | 75.30 | 23.50 | 271.25 | 400.26 | 0.68 |
|  |  |  | 37.36 | 0.61 | 0.61 | 1.78 | 1.00 | 3.64 | 5.89 | 0.00 |
|  | 2 | 4.370 |  |  |  |  |  |  |  |  |
|  |  |  | 9851.72 | -9.05 | -4.77 | 71.26 | 22.14 | 257.67 | 379.80 | 0.68 |
|  |  |  | 34.05 | 0.43 | 0.43 | 1.26 | 0.71 | 2.51 | 4.08 | 0.00 |
|  | 3 | 2.156 |  |  |  |  |  |  |  |  |
|  |  |  | 8574.55 | -13.67 | -10.01 | 62.16 | 18.90 | 229.24 | 336.02 | 0.68 |
|  |  |  | 32.26 | 0.43 | 0.43 | 1.26 | 0.70 | 2.42 | 3.94 | 0.00 |
|  | 4 | 2.632 |  |  |  |  |  |  |  |  |
|  |  |  | 1226.82 | -25.13 | -20.78 | 35.27 | 25.26 | 62.15 | 117.09 | 0.53 |
|  |  |  | 33.12 | 2.69 | 2.69 | 5.50 | 6.77 | 4.54 | 10.36 | 0.03 |
| 6598 | 1 | 1.044 | | | | | | | | |
| | | | 848.31 | -13.71 | -2.79 | 9.01 | 246.05 | 1.61 | 11.46 | 0.14 |
| | | | 18.65 | 1.49 | 1.49 | 2.37 | 54.87 | 0.13 | 1.15 | 0.01 |
| | 2 | 1.362 | | | | | | | | |
| | | | 867.00 | -10.76 | -3.03 | 7.87 | 139.76 | 1.74 | 10.56 | 0.16 |
| | | | 16.85 | 1.11 | 1.11 | 1.79 | 30.28 | 0.13 | 1.00 | 0.01 |
| | 3 | 0.942 | | | | | | | | |
| | | | 658.67 | -4.78 | -2.67 | 5.21 | 19.25 | 2.79 | 9.46 | 0.29 |
| | | | 17.40 | 0.71 | 0.71 | 1.22 | 6.07 | 0.29 | 1.21 | 0.02 |
| 6625 | 1 | 1.126 | | | | | | | | |
| | | | 897.06 | -8.97 | -4.61 | 10.95 | 39.36 | 5.96 | 20.02 | 0.30 |
| | | | 13.40 | 0.82 | 0.82 | 1.41 | 6.85 | 0.35 | 1.44 | 0.01 |
| | 2 | 1.332 | | | | | | | | |
| | | | 1117.42 | -8.18 | -3.69 | 9.47 | 44.88 | 4.34 | 16.17 | 0.27 |
| | | | 13.77 | 0.63 | 0.63 | 1.05 | 6.44 | 0.21 | 0.95 | 0.01 |
| | 3 | 0.950 | | | | | | | | |
| | | | 674.29 | -5.67 | -2.67 | 6.98 | 28.22 | 3.53 | 12.39 | 0.29 |
| | | | 15.51 | 0.86 | 0.86 | 1.45 | 7.81 | 0.32 | 1.39 | 0.02 |
| 6707 | 1 | 0.877 | | | | | | | | |
| | | | 506.94 | -4.87 | -2.05 | 8.91 | 22.46 | 6.12 | 18.02 | 0.34 |
| | | | 12.90 | 1.02 | 1.02 | 1.80 | 6.54 | 0.58 | 2.11 | 0.02 |
| | 2 | 1.189 | | | | | | | | |
| | | | 617.62 | -6.55 | -3.13 | 8.44 | 31.01 | 4.53 | 15.35 | 0.30 |
| | | | 12.69 | 0.89 | 0.89 | 1.52 | 7.53 | 0.36 | 1.52 | 0.02 |
| | 3 | 0.984 | | | | | | | | |
| | | | 537.87 | -9.95 | -4.69 | 8.34 | 64.45 | 2.87 | 12.83 | 0.22 |
| | | | 13.93 | 1.42 | 1.42 | 2.33 | 20.89 | 0.31 | 1.75 | 0.02 |
| 7087 | 1 | 0.943 | | | | | | | | |
| | | | 402.57 | -3.34 | -2.38 | 6.52 | 5.86 | 9.58 | 19.46 | 0.49 |
| | | | 13.34 | 0.65 | 0.65 | 1.30 | 1.95 | 0.95 | 2.31 | 0.03 |
| | 2 | 1.013 | | | | | | | | |
| | | | 593.73 | -2.42 | -1.71 | 4.95 | 4.24 | 7.56 | 15.11 | 0.50 |
| | | | 13.39 | 0.32 | 0.32 | 0.65 | 0.93 | 0.47 | 1.14 | 0.02 |
| | 3 | 1.033 | | | | | | | | |
| | | | 672.98 | -2.22 | -1.74 | 3.95 | 2.76 | 7.12 | 13.29 | 0.54 |
| | | | 13.82 | 0.21 | 0.21 | 0.45 | 0.54 | 0.37 | 0.85 | 0.02 |
| | 4 | 1.027 | | | | | | | | |
| | | | 71.67 | -3.62 | -3.15 | 4.12 | 2.70 | 7.82 | 14.31 | 0.55 |
| | | | 9.17 | 1.43 | 1.43 | 3.03 | 3.43 | 2.72 | 5.98 | 0.13 |
| 7156 | 1 | 0.986 | | | | | | | | |

|  |  |  |  |  |  |  |  |  |  |
|---|---|---|---|---|---|---|---|---|---|
|  |  |  | 496.78 | -6.91 | -2.49 | 9.95 | 42.41 | 4.86 | 17.41 | 0.28 |
|  |  |  | 16.58 | 1.83 | 1.83 | 3.11 | 17.38 | 0.68 | 2.99 | 0.03 |
|  | 2 | 1.115 |  |  |  |  |  |  |  |  |
|  |  |  | 858.22 | -7.46 | -2.93 | 7.98 | 51.17 | 3.06 | 12.74 | 0.24 |
|  |  |  | 17.32 | 0.99 | 0.99 | 1.65 | 12.77 | 0.26 | 1.33 | 0.01 |
|  | 3 | 1.538 |  |  |  |  |  |  |  |  |
|  |  |  | 1038.11 | -5.62 | -3.09 | 5.61 | 24.51 | 2.71 | 9.77 | 0.28 |
|  |  |  | 22.70 | 0.69 | 0.69 | 1.15 | 6.59 | 0.24 | 1.07 | 0.02 |
|  | 4 | 0.698 |  |  |  |  |  |  |  |  |
|  |  |  | 143.43 | -1.56 | -1.51 | 0.62 | 0.27 | 1.67 | 2.70 | 0.62 |
|  |  |  | 49.19 | 1.00 | 1.00 | 1.72 | 1.37 | 1.11 | 2.78 | 0.49 |
| 7293 | 1 | 0.962 |  |  |  |  |  |  |  |  |
|  |  |  | 633.11 | -0.94 | -0.57 | 5.35 | 1.95 | 16.79 | 25.81 | 0.65 |
|  |  |  | 14.65 | 0.26 | 0.26 | 0.67 | 0.44 | 1.00 | 1.77 | 0.02 |
|  | 2 | 0.933 |  |  |  |  |  |  |  |  |
|  |  |  | 1479.32 | -1.10 | -0.56 | 5.03 | 3.05 | 10.19 | 18.19 | 0.56 |
|  |  |  | 16.44 | 0.13 | 0.13 | 0.30 | 0.31 | 0.28 | 0.61 | 0.01 |
|  | 3 | 1.696 |  |  |  |  |  |  |  |  |
|  |  |  | 2349.31 | -1.47 | -0.95 | 3.67 | 3.04 | 5.76 | 11.38 | 0.51 |
|  |  |  | 21.19 | 0.08 | 0.08 | 0.17 | 0.24 | 0.12 | 0.30 | 0.01 |
|  | 4 | 0.995 |  |  |  |  |  |  |  |  |
|  |  |  | 241.42 | -1.62 | -1.32 | 2.00 | 1.75 | 2.99 | 6.03 | 0.50 |
|  |  |  | 15.27 | 0.22 | 0.22 | 0.38 | 0.55 | 0.34 | 0.78 | 0.03 |
| 7403 | 1 | 0.958 |  |  |  |  |  |  |  |  |
|  |  |  | 388.72 | -2.97 | -1.99 | 10.19 | 5.42 | 23.02 | 39.48 | 0.58 |
|  |  |  | 10.79 | 0.64 | 0.64 | 1.50 | 1.39 | 1.83 | 3.59 | 0.03 |
|  | 2 | 0.996 |  |  |  |  |  |  |  |  |
|  |  |  | 686.40 | -2.41 | -1.68 | 6.61 | 4.12 | 13.08 | 23.56 | 0.56 |
|  |  |  | 11.15 | 0.27 | 0.27 | 0.59 | 0.64 | 0.57 | 1.22 | 0.02 |
|  | 3 | 0.978 |  |  |  |  |  |  |  |  |
|  |  |  | 793.94 | -1.45 | -1.17 | 3.06 | 1.51 | 7.38 | 12.37 | 0.60 |
|  |  |  | 15.77 | 0.14 | 0.14 | 0.33 | 0.29 | 0.34 | 0.69 | 0.02 |
|  | 4 | 0.809 |  |  |  |  |  |  |  |  |
|  |  |  | 103.66 | -2.42 | -2.14 | 1.63 | 1.79 | 2.04 | 4.45 | 0.46 |
|  |  |  | 18.53 | 0.94 | 0.94 | 1.86 | 3.34 | 1.09 | 2.92 | 0.18 |
| 7588 | 1 | 1.190 |  |  |  |  |  |  |  |  |
|  |  |  | 1122.41 | -1.75 | -0.66 | 4.35 | 7.81 | 3.79 | 9.85 | 0.39 |
|  |  |  | 19.67 | 0.29 | 0.29 | 0.52 | 1.43 | 0.21 | 0.69 | 0.02 |
|  | 2 | 1.161 |  |  |  |  |  |  |  |  |
|  |  |  | 1418.63 | -1.57 | -0.87 | 3.57 | 4.60 | 3.96 | 9.12 | 0.43 |
|  |  |  | 20.08 | 0.16 | 0.16 | 0.30 | 0.63 | 0.16 | 0.45 | 0.01 |

|  |  |  |  |  |  |  |  |  |  |
|---|---|---|---|---|---|---|---|---|---|
|  | 3 | 0.802 |  |  |  |  |  |  |  |
|  |  |  | 1042.00 | -1.05 | -0.79 | 2.12 | 1.48 | 3.80 | 7.11 | 0.53 |
|  |  |  | 22.14 | 0.11 | 0.11 | 0.24 | 0.29 | 0.20 | 0.46 | 0.02 |
|  | 4 | 0.865 |  |  |  |  |  |  |  |
|  |  |  | 78.72 | -2.21 | -1.96 | 1.90 | 1.47 | 3.15 | 6.08 | 0.52 |
|  |  |  | 18.24 | 1.28 | 1.28 | 2.42 | 3.20 | 1.88 | 4.42 | 0.21 |
| 7648 | 1 | 1.091 |  |  |  |  |  |  |  |
|  |  |  | 583.12 | -2.12 | -0.78 | 6.29 | 9.06 | 6.42 | 15.40 | 0.42 |
|  |  |  | 15.78 | 0.62 | 0.62 | 1.15 | 2.60 | 0.58 | 1.68 | 0.03 |
|  | 2 | 0.940 |  |  |  |  |  |  |  |
|  |  |  | 761.60 | -3.71 | -2.00 | 6.84 | 12.31 | 5.93 | 15.43 | 0.38 |
|  |  |  | 13.55 | 0.46 | 0.46 | 0.84 | 2.31 | 0.35 | 1.12 | 0.02 |
|  | 3 | 0.925 |  |  |  |  |  |  |  |
|  |  |  | 984.16 | -4.18 | -2.47 | 5.77 | 13.25 | 4.23 | 12.02 | 0.35 |
|  |  |  | 14.83 | 0.36 | 0.36 | 0.63 | 2.12 | 0.21 | 0.75 | 0.01 |
|  | 4 | 0.809 |  |  |  |  |  |  |  |
|  |  |  | 99.92 | -3.57 | -2.47 | 3.10 | 9.26 | 1.91 | 5.98 | 0.32 |
|  |  |  | 13.18 | 2.03 | 2.03 | 3.38 | 14.11 | 0.97 | 3.73 | 0.12 |
| 8049 | 1 | 1.071 |  |  |  |  |  |  |  |
|  |  |  | 517.18 | -16.93 | -4.24 | 15.63 | 191.91 | 4.18 | 22.16 | 0.19 |
|  |  |  | 12.88 | 2.64 | 2.64 | 4.66 | 57.83 | 0.55 | 3.40 | 0.01 |
|  |  |  | 679.07 | 15.63 | 20.29 | 13.75 | 38.40 | 8.82 | 26.97 | 0.33 |
|  |  |  | 14.63 | 1.24 | 1.24 | 2.03 | 8.05 | 0.61 | 2.28 | 0.02 |
|  |  |  | 186.45 | 54.69 | 63.33 | 15.38 | 96.88 | 5.95 | 24.64 | 0.24 |
|  |  |  | 10.48 | 5.43 | 5.43 | 8.95 | 68.12 | 1.48 | 7.50 | 0.04 |
|  | 2 | 0.961 |  |  |  |  |  |  |  |
|  |  |  | 865.67 | -8.47 | -2.89 | 11.30 | 57.31 | 4.97 | 18.97 | 0.26 |
|  |  |  | 12.64 | 0.46 | 0.46 | 0.80 | 4.95 | 0.20 | 0.86 | 0.01 |
|  |  |  | 926.20 | 16.08 | 20.36 | 11.77 | 36.63 | 7.03 | 22.38 | 0.31 |
|  |  |  | 12.10 | 0.44 | 0.44 | 0.73 | 3.11 | 0.22 | 0.83 | 0.01 |
|  |  |  | 98.19 | 19.81 | 56.21 | 21.07 | 1247.02 | 2.65 | 24.82 | 0.11 |
|  |  |  | 8.82 | 5.46 | 5.46 | 8.13 | 294.48 | 0.41 | 4.24 | 0.01 |
|  | 3 | 1.132 |  |  |  |  |  |  |  |
|  |  |  | 907.31 | -6.43 | -3.09 | 8.65 | 29.58 | 4.87 | 16.04 | 0.30 |
|  |  |  | 12.56 | 0.61 | 0.61 | 1.07 | 5.01 | 0.28 | 1.14 | 0.01 |
|  |  |  | 1021.02 | 17.33 | 20.46 | 8.72 | 26.61 | 5.28 | 16.68 | 0.32 |
|  |  |  | 12.81 | 0.51 | 0.51 | 0.85 | 3.65 | 0.23 | 0.90 | 0.01 |
|  | 4 | 0.996 |  |  |  |  |  |  |  |
|  |  |  | 49.61 | -2.33 | -2.29 | 1.17 | 0.20 | 7.24 | 9.35 | 0.77 |

| | | | | | | | |
|---|---|---|---|---|---|---|---|
| 17.34 | 3.67 | 3.67 | 5.54 | 1.84 | 4.31 | 10.23 | 0.71 |
| 85.40 | 19.11 | 20.85 | 4.75 | 15.02 | 2.81 | 9.00 | 0.31 |
| 10.45 | 2.99 | 2.99 | 5.05 | 22.24 | 1.32 | 5.28 | 0.11 |

TABLE 3. Spectral Fit Parameters

| Trigger | Interval [a] | $\chi^2/\nu$ | A [b] $\varepsilon_A$ | $\alpha$ $\varepsilon_\alpha$ | $\beta$ $\varepsilon_\beta$ | $E_{peak}$ [c] $\varepsilon_{E_{peak}}$ |
|---|---|---|---|---|---|---|
| 1406 | [ -0.704, 5.952] | 0.288 | 0.044 0.009 | -1.203 0.149 | -2.897 0.214 | 79.434 21.396 |
|  | [ 5.952, 17.216] | 1.892 | 0.204 0.042 | -0.485 0.150 | -2.215 0.025 | 97.701 16.791 |
| 2193 | [ -1.024, 23.040] | 2.253 | 0.098 0.002 | -0.576 0.034 | -3.196 0.213 | 378.794 21.954 |
|  | [ 23.040, 75.776] | 2.725 | 0.576 0.024 | 0.023 0.039 | -2.581 0.043 | 122.812 3.811 |
| 2387 | [ 0.256, 9.984] | 0.937 | 0.439 0.045 | -0.026 0.085 | -2.463 0.033 | 106.791 8.001 |
|  | [ 9.984, 26.880] | 1.164 | 0.598 0.051 | -0.321 0.068 | -2.606 0.025 | 107.401 6.596 |
| 3256 | [ -0.772, 14.008] | 0.184 | 0.268 0.068 | 0.245 0.212 | -2.948 0.131 | 110.537 16.512 |
| 3257 | [ -1.280, 9.984] | 0.767 | 0.142 0.013 | -0.241 0.093 | -2.790 0.120 | 169.240 16.323 |
|  | [ 9.984, 35.072] | 1.464 | 0.373 0.006 | -0.164 0.021 | -3.046 0.106 | 122.459 1.834 |
| 6147 | [ -1.024, 6.144] | 0.072 | 0.044 0.012 | -0.923 0.228 | -3.801 1.284 | 76.394 22.327 |
|  | [ 6.144, 17.408] | 0.415 | 0.204 0.041 | -0.362 0.168 | -3.892 0.481 | 79.412 11.468 |
| 6504 | [ 0.000, 6.656] | 0.197 | 0.190 0.059 | 0.588 0.263 | -2.784 0.166 | 120.200 20.339 |
|  | [ 6.656, 20.992] | 0.582 | 0.617 0.125 | 0.668 0.166 | -2.591 0.058 | 110.757 11.193 |
| 6625 | [ -1.024, 7.168] | 0.290 | 0.112 0.018 | -1.176 0.116 | -4.540 1.761 | 53.802 9.309 |
|  | [ 7.168, 18.944] | 0.339 | 0.276 0.036 | -0.717 0.099 | -3.386 0.188 | 78.318 8.623 |
| 7293 | [ -0.000, 7.680] | 0.318 | 0.117 0.015 | -0.149 0.124 | -2.809 0.108 | 151.082 17.411 |
|  | [ 7.680, 22.528] | 0.261 | 0.791 | 0.754 | -2.824 | 116.054 |

|      |                    |       | 0.117 | 0.132  | 0.050  | 8.783   |
|------|--------------------|-------|-------|--------|--------|---------|
| 7588 | [ -0.768,  4.352]  | 0.186 | 0.047 | -1.732 | -2.804 | 12.072  |
|      |                    |       | 0.030 | 0.382  | 0.079  | 17.910  |
|      | [ 4.352, 13.568]   | 0.435 | 0.135 | -1.192 | -2.959 | 65.191  |
|      |                    |       | 0.017 | 0.096  | 0.106  | 10.520  |
| 7648 | [ -1.536,  6.144]  | 0.214 | 0.027 | -0.782 | -2.427 | 146.933 |
|      |                    |       | 0.009 | 0.264  | 0.207  | 58.004  |
|      | [ 6.144, 18.432]   | 0.445 | 0.122 | -0.292 | -2.571 | 149.529 |
|      |                    |       | 0.015 | 0.114  | 0.088  | 17.817  |
| 8049 | [ -1.024,  8.704]  | 0.651 | 0.050 | -0.491 | -2.567 | 138.476 |
|      |                    |       | 0.012 | 0.229  | 0.170  | 37.291  |
|      | [ 8.704, 18.944]   | 1.194 | 0.136 | -0.476 | -3.033 | 117.874 |
|      |                    |       | 0.016 | 0.103  | 0.133  | 12.583  |
|      | [ 20.992, 33.280]  | 0.757 | 0.051 | -1.468 | -2.632 | 119.059 |
|      |                    |       | 0.005 | 0.083  | 0.157  | 28.601  |
|      | [ 33.280, 50.688]  | 1.417 | 0.123 | -1.213 | -2.922 | 91.976  |
|      |                    |       | 0.005 | 0.038  | 0.108  | 6.235   |

[a] Spectral intervals (s) relative to burst trigger time
[b] Amplitude in photons cm$^{-2}$ s$^{-1}$ keV$^{-1}$
[c] Energy in keV

TABLE 4. $\tau_1$ and $\tau_2$ Ranges [a]

| $\tau_1$ | $\tau_2$ | $\mu$ [b] | $(1 + 4\mu)$ |
|---|---|---|---|
| 1.9 → 250. | 2.3 → 26. | 10. → 0.3 | 43. → 2.1 |
| 2.7 → 57. | 2.8 → 17. | 4.5 → 0.4 | 19. → 2.6 |
| 1.5 → 28. | 2.6 → 9. | 3.3 → 0.4 | 14. → 2.6 |

[a] Approximate ±1 σ ranges
[b] $\mu_{max} = (\tau_{1max}/\tau_{2min})^{1/2}$, $\mu_{min} = (\tau_{1min}/\tau_{2max})^{1/2}$

TABLE 5. $\tau_{peak} = \exp\{(\kappa - \kappa_0)/p\}$ Relation

| Channel | $\kappa_0$ | p |
|---|---|---|
| 1 | 0.79 | –0.42 |
| 2 | 0.68 | –0.34 |
| 3 | 0.61 | –0.30 |

TABLE 6. $w/w_0 = \tau_{peak}^{p}$ Relation

| Channel | $w_0$ | p |
|---|---|---|
| 1 | 1.02 | 0.71 |
| 2 | 0.54 | 0.91 |
| 3 | 0.38 | 1.02 |

APPENDIX 1

Here we detail the partial derivatives, the expressions for error propagation for our adopted pulse model, and its integral. We also describe a convenient approach for conversion of automated measurements or manual inputs into initial guesses for parameter values for the purposes of curve fitting. The pulse model is a form proportional to the inverse of the product of two exponentials, one increasing and one decreasing with time,

$$I(t) = A \lambda / [\exp\{\tau_1/t\} \exp\{t/\tau_2\}] = A \lambda / \exp\{\tau_1/t + t/\tau_2\} \quad \text{for } t > 0. \tag{A1}$$

It is convenient to define auxiliary parameters: $\mu = (\tau_1/\tau_2)^{1/2}$, $\lambda = \exp(2\mu)$, and $B = \tau_1/t + t/\tau_2$. Also, the model intensity is normalized by $\lambda$ so that at $t = \tau_{peak} = (\tau_1\tau_2)^{1/2}$, the intensity equals the peak amplitude, $A$. In addition to $A$, $\tau_1$, $\tau_2$, the pulse position is defined by the start time, $t_s$, relative to some (arbitrary) initial point of the time series. For an expeditious curve fitting program we need partial derivatives with respect to these four parameters:

$$\partial I / \partial A = I / A, \tag{A2}$$

$$\partial I / \partial t_s = A \lambda \exp\{-B\} [-\tau_1/t^2 + 1/\tau_2], \tag{A3}$$

$$\partial I / \partial \tau_1 = A \lambda \exp\{-B\} [(\tau_1\tau_2)^{-1/2} - 1/t], \text{ and} \tag{A4}$$

$$\partial I / \partial \tau_2 = A \lambda \exp\{-B\} [t/\tau_2^2 - \mu/\tau_2]. \tag{A5}$$

The pulse width from eq. (2), $w = \tau_2 (1 + 4\mu)^{1/2}$, asymmetry from eq. (4), $\kappa = (1 + 4\mu)^{-1/2}$, and peak time involve only the characteristic timescales, $\tau_1$ and $\tau_2$. We must propagate the formal fitted variances of the latter two parameters to arrive at the total variances for the derived pulse shape and position parameters. The total variance in the width parameter can be expressed as

$$\sigma_w^2 = (1 + 4\mu)\sigma_{\tau 2}^2 + \tau_2^2[\mu^2/(1 + 4\mu)]\{(\sigma_{\tau 1}/\tau_1)^2 + (\sigma_{\tau 2}/\tau_2)^2\}, \tag{A6}$$

and for the asymmetry parameter,

$$\sigma_\kappa^2 = [\mu^2/(1 + 4\mu)^3]\{(\sigma_{\tau 1}/\tau_1)^2 + (\sigma_{\tau 2}/\tau_2)^2\}. \tag{A7}$$

The variance for the peak time is

$$\sigma_{\tau peak}^2 = \tfrac{1}{4} \tau_{peak}^{-2} \{(\tau_2\sigma_{\tau 1})^2 + (\tau_1\sigma_{\tau 2})^2\}. \tag{A8}$$

The total variance for a peak lag measurement is relatively large since it involves propagation of the variances for the peaks in the two lagged energy channels as well as for the two start times:

$$\sigma_{peaklag}^2 = \sigma_{\tau peak1}^2 + \sigma_{ts1}^2 + \sigma_{\tau peak2}^2 + \sigma_{ts2}^2 \ . \tag{A9}$$

An analogous formulation obtains for the variance of a pulse centroid lag measurement, but with the $\sigma_{\tau peak}$ values replaced by errors on the time of pulse centroid, $\sigma_{tcen}$. Values for $\sigma_{tcen}$ were estimated by computing a pulse's first moment, $t_{cen} = \int I(t) \, t \, dt / \int I(t) \, dt$, with the actual counts and with random Poisson variance added (including that of the background level), and then differencing. The variance for a centroid lag measurement is then

$$\sigma_{cenlag}^2 = \sigma_{tcen2}^2 + \sigma_{ts1}^2 + \sigma_{tcen2}^2 + \sigma_{ts2}^2 \ , \tag{A10}$$

which is dominated by the $\sigma_{ts}^2$ terms since centroid times are much better determined than are the times of pulse peaks.

The automated procedure for generating first estimates for parameter values for pulse fitting utilizes a Bayesian Block representation of the time profile. Per energy channel the significant peaks in the time profile are selected, requiring that retained peaks are > 0.05 times the maximum peak amplitude (usually, just one or two peaks were identified by this process). Valleys between "significant" adjacent peaks are required to be at least 10% lower than the less intense of the two peaks. Amplitude estimates are just the intensities of the peaks, unadjusted for pulse overlap. The $t_s$ estimates are the beginning of the first Bayesian block after the valley preceding a peak. In the case of the first pulse, the first valley is defined as the first block with intensity at > 0.05 of the most intense peak. It suffices to set the $\tau_1$ and $\tau_2$ estimates as just the time of peak – $t_s$. A branch in the fitting program for manual intervention can be selected, to facilitate convergence or to alter the number of pulses determined by the initial, automated approach. The number of pulses is entered. For each pulse, three points are picked interactively, resulting in the times and intensities at 1/e amplitude on the rise and decay sides of the pulse – $t_{on}$ and $t_{off}$ respectively – and at the peak itself. The amplitude guess is then the intensity of the chosen peak. Relative to $t_{on}$, define $\tau_{peak}$ = time of selected peak – $t_{on}$, and $\Delta t_{1/e} = t_{off} - t_{on}$. The remaining first guesses for the parameter values are $t_s = t_{on}$, and from eq. (A1),

$$\tau_1 = \ln(10) \, [1/\Delta t_{1/e} + \Delta t_{1/e}/\tau_{peak}^2 - 2/\tau_{peak}]^{-1}, \text{ and} \tag{A11}$$

$$\tau_2 = [\Delta t_{1/e} + \tau_{peak}^2/\Delta t_{1/e} - 2\tau_{peak}] / \ln(10) \ . \tag{A12}$$

The integral of eq. (A1) over the pulse duration yields the total counts within a model pulse,

$$S = \int_0^\infty I(t)\, dt = A\, \lambda \int_0^\infty \exp\{-\tau_1/t - t/\tau_2\}\, dt \qquad (A13)$$

This integral is given as 3.324.1 (p. 307) in Gradshteyn and Ryzhik's Table of Integrals, Series, and Products (2000) and leads to the result that

$$S = A\, \lambda\, (4\tau_1\tau_2)^{1/2}\, K_1([4\tau_1/\tau_2]^{1/2}) = 2\, A\, \lambda\, \tau_{peak}\, K_1(2\mu) \qquad (A14)$$

where $K_1(x)$ is a Bessel function of imaginary argument.

## APPENDIX 2

The following figures illustrate the temporal fits obtained using the model of eq. (1) in the four energy channels for four representative bursts from Table 2. Each time profile is background subtracted, shown as a histogram, with a solid line for the total model fit, and long dashed lines for the individual pulse fits. These examples include single, double, and triple pulse bursts, in which some of the pulses are well defined through channel 4 while others diminish to negligible intensity. The general trend of decreasing width with energy (see Figures 5 and 6) is evident. The occasional inadequacy of the pulse model in accommodating low-level emission in the pulse tail (Figure 24) or unusually spiky emission near the pulse peak (Figure 25) is shown. The last example (Figure 26) illustrates a triple-pulse burst where pulse overlap is substantial in channel 1, diminishing as pulses narrow in the higher energy channels. Overall, the pulse model affords satisfactory fits for the rise, peak, and decay portions of nearly all pulses in Table 2.

Figure Captions

Fig. 1–Pulse model shape parameters $\tau_2$ vs. $\tau_1$ for pulses in energy channels 1 (a), 2 (b), and 3 (c). In the log-log plot for channel 1, the dashed line indicates the best linear fit to the tentative trend for, $\log(\tau_2) = -0.28 \times \log(\tau_1) + 1.22$. The substantial outlier, the single pulse burst trigger 6526, is omitted from the plot and the fit. For channel 2 (3), the slope of the dashed line is -0.21 (0.08).

Fig. 2–Pulse asymmetry vs. width for pulses in energy channels 1 (a), 2 (b), and 3 (c). The single pulse burst trigger 6526 is the outlier with width > 100 s.

Fig. 3–Pulse asymmetry vs. $\tau_{peak}$ (peak time) for pulses in energy channels 1 (a), 2 (b), and 3 (c). In the log-log plot for channel 1, the dashed line indicates the best linear fit to the tentative trend, with a slope of -0.42. For channel 2 (3), the slope of the dashed line is 0.34 (0.30).

Fig. 4–Pulse $\tau_{peak}$ (peak time) vs. width for pulses in energy channels 1 (a), 2 (b), and 3 (c). In the log-log plot for channel 1, the dashed line indicates the best linear fit to the tentative trend, with a slope of 0.71. For channel 2 (3), the slope of the dashed line is 0.91 (1.02).

Fig. 5–Energy channel vs. pulse width. Symbols joined by line segments correspond to the same pulse. Within a pulse, the general trend of increasing pulse width with decreasing energy is apparent.

Fig. 6–Pulse width histograms for four energy channels. In all four panels the outlier bin > 100 s is the single pulse burst trigger 6526. The shift of the mode to larger widths at lower energy channels is apparent.

Fig. 7–Pulse peak lag vs. width. The width parameter for energy channel 2 is plotted. Peak lags were determined between energy channels 2 and 3. A general increase of pulse width with spectral lag is apparent.

Fig. 8–Pulse centroid lag vs. width. In panel a, the width parameter for energy channel is plotted, with centroid lags determined between energy channels 1 and 3. In the log-log plot, the dashed line indicates the best linear fit with a slope of 1.42. In panel b, the width is for channel 2, and centroid lags are between energy channels 2 and 4 (fewer pulses are available due to frequent absence of usable signal in channel 4).

Fig. 9–Alpha spectral parameter histogram. The low energy power-law index, $\alpha$, is from Band function fits to 25 spectra for rising and decaying intervals within individual pulses. The mode is the bin $\alpha = -0.5$ to $-0.2$.

Fig. 10–As in Figure 9, for beta spectral parameter. The mode is the bin $\beta = -2.9$ to $-2.6$.

Fig. 11–As in Figure 9, for peak energy (in $\nu F[\nu]$) spectral parameter. The distribution is approximately symmetrical about $E_{pk} \sim 110$ keV. The notable outlier near 400 keV is the rising interval for the single pulse burst trigger 2193.

Fig. 12–Hardness ratio (3/1) histograms. The fluence hardness ratios comparing energy channel 3 and energy channel 1 are shown for a sample of 93 long-lag bursts (solid histogram) and 714 short-lag bursts (dotted histogram). The area of the short-lag distribution is normalized to the area of the long-lag distribution.

Fig. 13–As in Figure 12, for hardness ratio (3/2).

Fig. 14–Spectral shape parameters alpha vs. beta. The lower energy ($\alpha$) and high energy ($\beta$) power-law indices from Band function fits to 25 spectra for intervals within individual pulses are plotted. No correlation is apparent.

Fig. 15–Spectral shape parameters low energy ($\alpha$) power-law index vs. peak energy ($E_{pk}$). No correlation is apparent.

Fig. 16–Spectral shape parameters high energy ($\beta$) power-law index vs. peak energy ($E_{pk}$). No correlation is apparent.

Fig. 17–Spectral shape parameter alpha vs. pulse width. The low energy Band function ($\alpha$) power-law indices for the intervals within each pulse are plotted against pulse width. Two intervals are plotted for each pulse; a diamond symbol for the first interval, a square symbol for the second interval.

Fig. 18–As in Figure 17, for spectral shape parameter beta vs. pulse width.

Fig. 19–As in Figure 17, for spectral shape parameter peak energy vs. pulse width.

Fig. 20a–Spectral shape parameters alpha vs. pulse asymmetry. The low energy Band function ($\alpha$) power-law indices for intervals within individual pulses are plotted against pulse asymmetry. Symbols have same meaning as in Figure 17. A trend of flatter $\alpha$ corresponding to higher pulse asymmetry is suggested.

Fig. 20b–As in Fig. 20a for $\alpha$ averaged over a single pulse. The best linear fit to the tentative trend is shown as a dashed line with a slope of 2.58.

Fig. 21–As in Fig. 20a, for spectral shape parameter beta vs. pulse asymmetry.

Fig. 22–As in Fig. 20a, for spectral shape parameter peak energy vs. pulse asymmetry.

Fig. 23–GRB 930214c, a two-pulse burst in which the second pulse diminishes in intensity relative to the first pulse, becoming negligible in channel 4. The main pulse decreases in width across channels 1$\rightarrow$4 as 82, 55, 33, and 18 seconds, with associated asymmetry parameter values of 0.61, 0.48, 0.44, and 0.44.

Fig. 24–GRB 930612a, a single-pulse burst with evidence of additional low-level emission (not modeled) in the pulse tail in channels 1 and 2. The pulse decreases in width across channels 1$\rightarrow$4 as 24, 20, 15, and 12 seconds, with associated asymmetry parameter values of 0.50, 0.43, 0.37, and 0.38.

Fig. 25–GRB 990102a, a single-pulse burst in which emission near the peak is spikier than accommodated by the pulse model in channels 1, 2 and 3. The pulse decreases in width across channels 1$\rightarrow$4 as 26, 18, 11 and 6 seconds, with associated asymmetry parameter values of 0.65, 0.56, 0.51, and 0.50.

Fig. 26–GRB 000323, a three-pulse burst with substantial overlap between the first two pulses in channel 1. The third pulse decreases in intensity with energy, becoming negligible in channel 4, where the first two pulses as well are not well defined. The first pulse decreases in width across channels 1$\rightarrow$3 as 22, 19, and 16 seconds, with associated asymmetry parameter values of 0.19, 0.26, and 0.30; the second pulse decreases in width as 27, 22, and 17 seconds, with asymmetries of 0.33, 0.31, and 0.32.

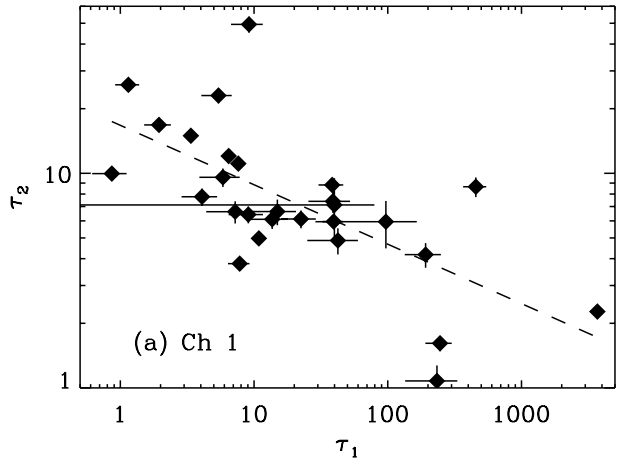

Fig1a_Tau1-vs-Tau2_c1.eps

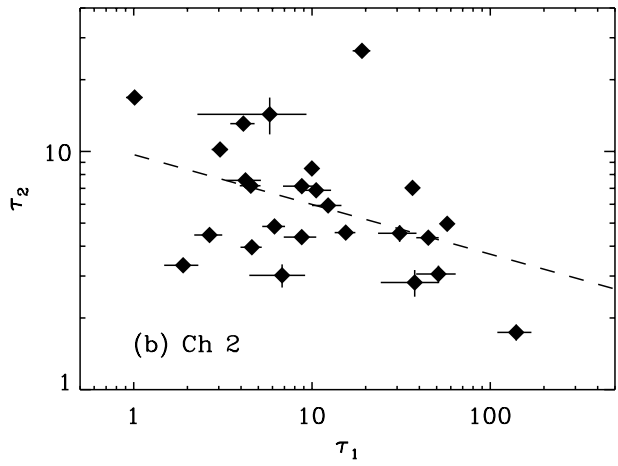

Fig1b_Tau1-vs-Tau2_c2.eps

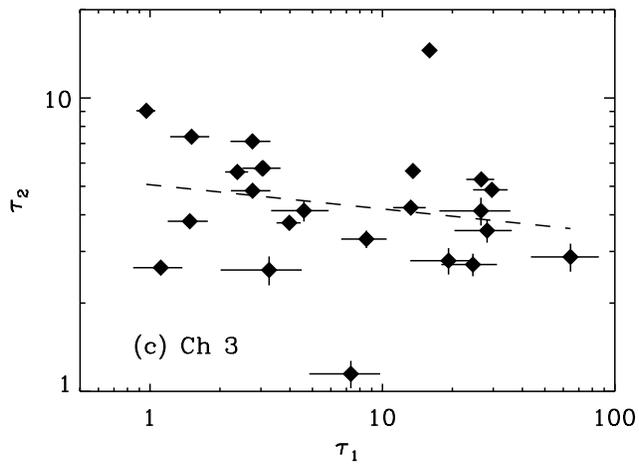

Fig1c_Tau1-vs-Tau2_c3.eps

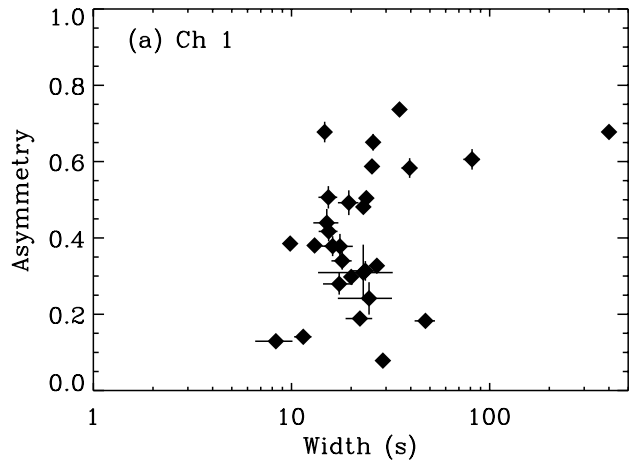

Fig2a_Width-vs-Asymmetry_c1.eps

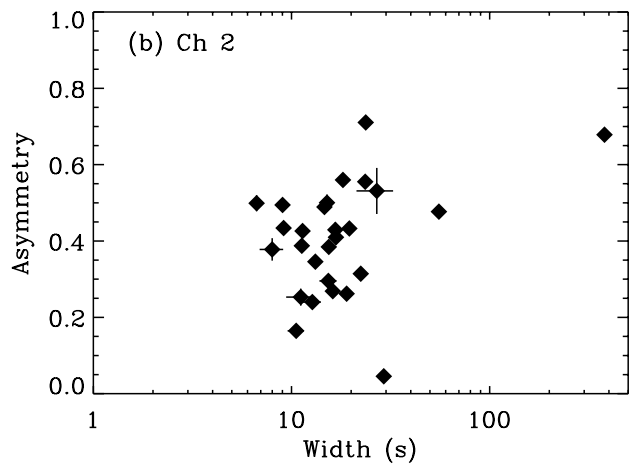

Fig2b_Width-vs-Asymmetry_c2.eps

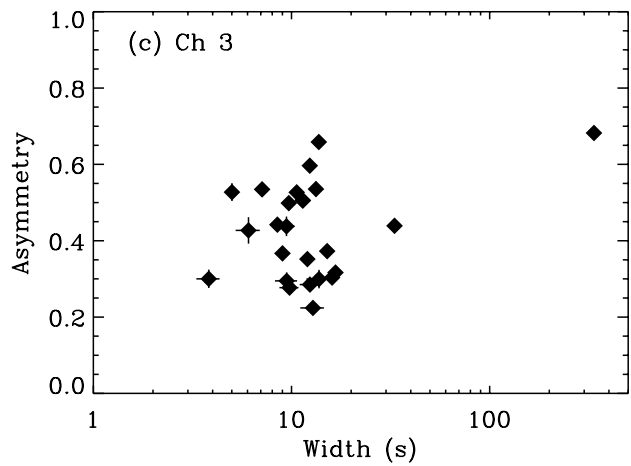

Fig2c_Width-vs-Asymmetry_c3.eps

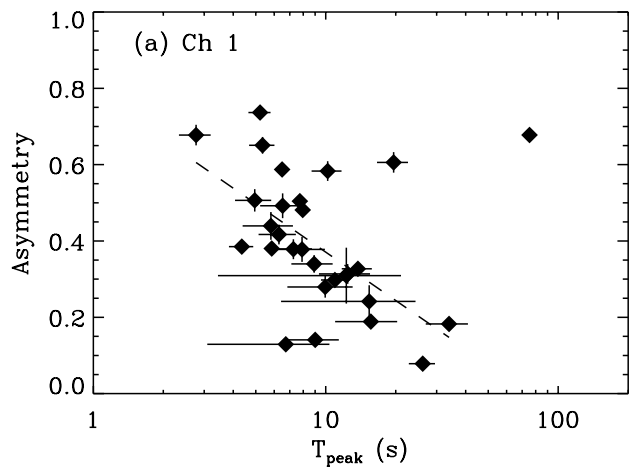

Fig3a_Tpeak-vs-Asymmetry_c1.eps

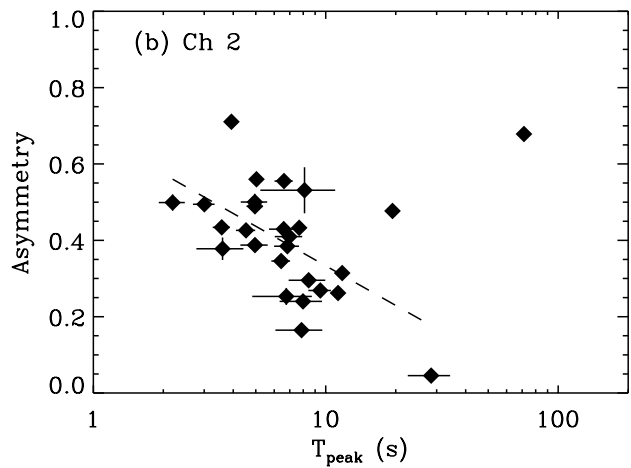

Fig3b_Tpeak-vs-Asymmetry_c2.eps

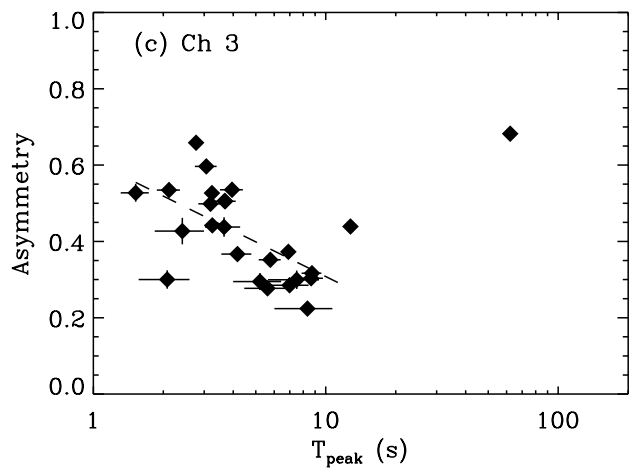

Fig3c_Tpeak-vs-Asymmetry_c3.eps

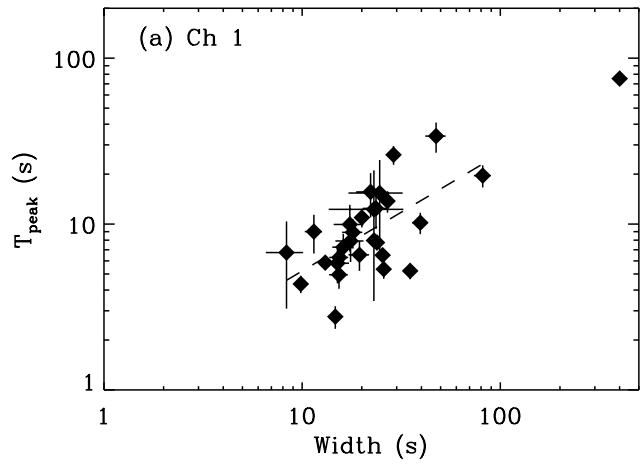

Fig4a_Width-vs-Tpeak_c1.eps

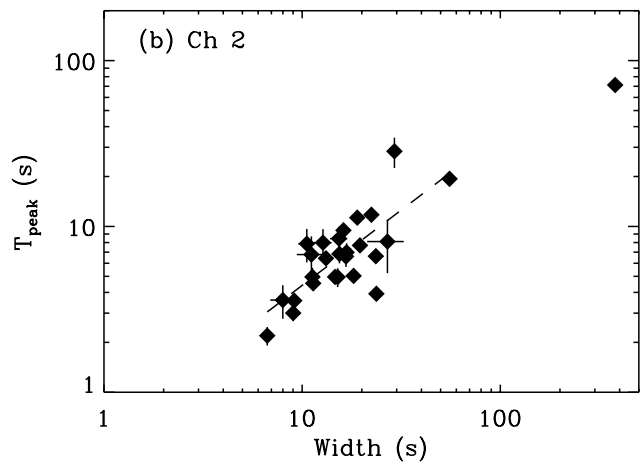

Fig4b_Width-vs-Tpeak_c2.eps

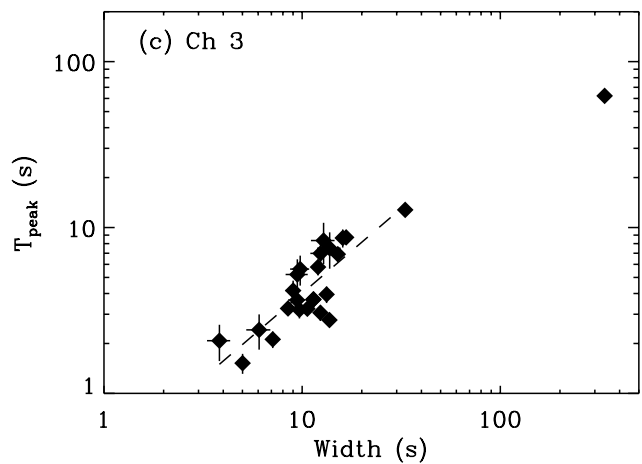

Fig4c_Width-vs-Tpeak_c3.eps

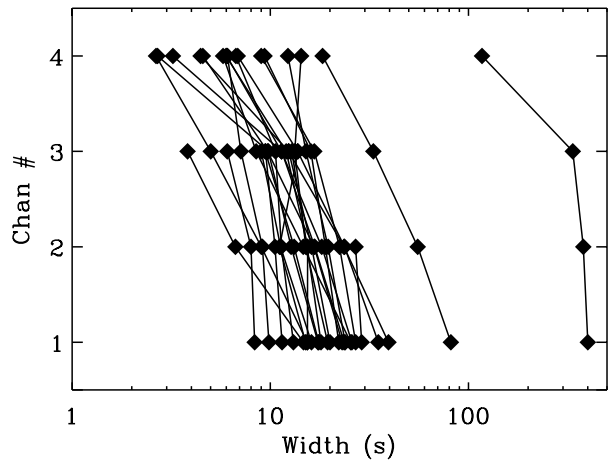

Fig05_Width-vs-Chan#.eps

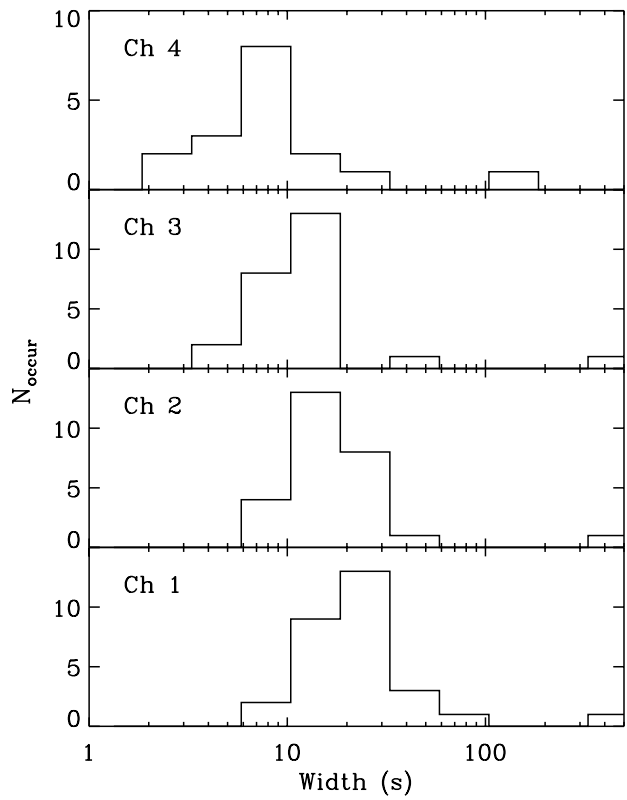

Fig06_Width-histos.eps

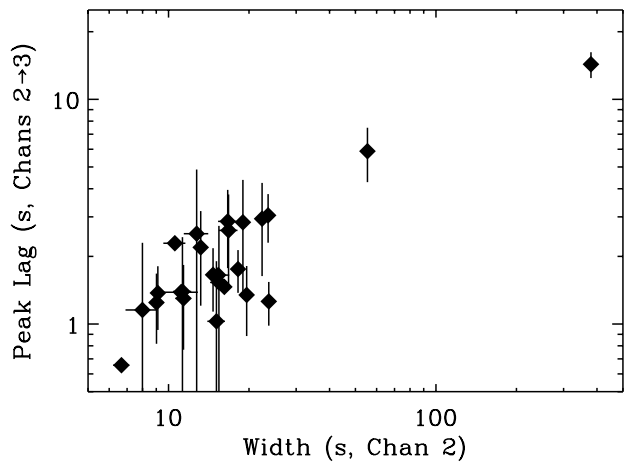

Fig07_Width-vs-PeakLag.eps

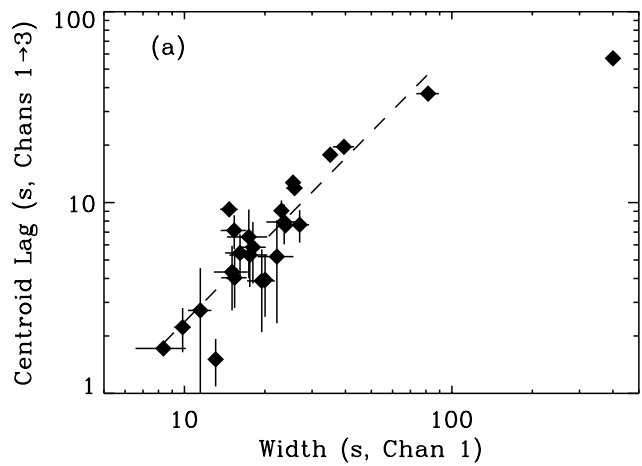

Fig8a_Width-vs-CentroidLag13.eps

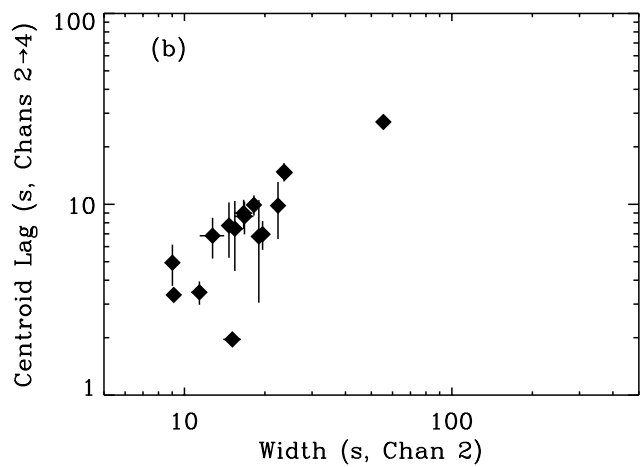

Fig8b_Width-vs-CentroidLag24.eps

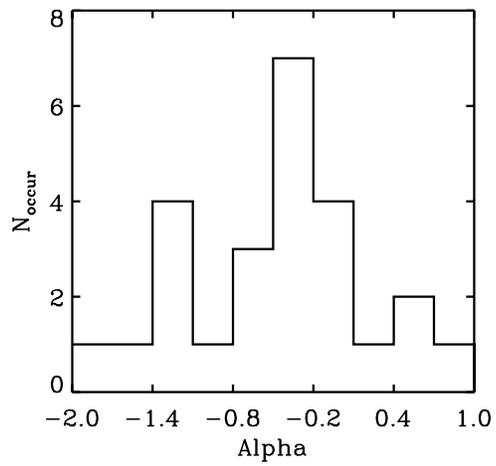

Fig09_Alpha-histo.eps

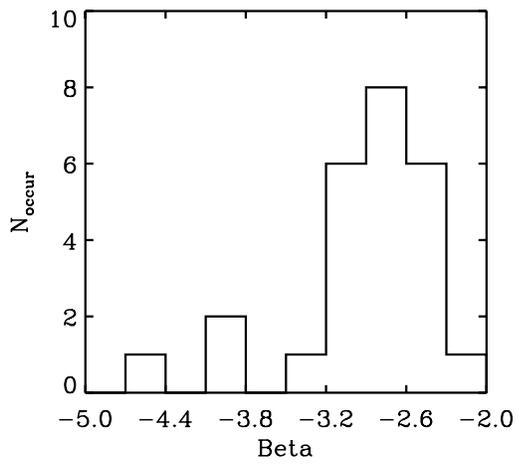

Fig10_Beta-histo.eps

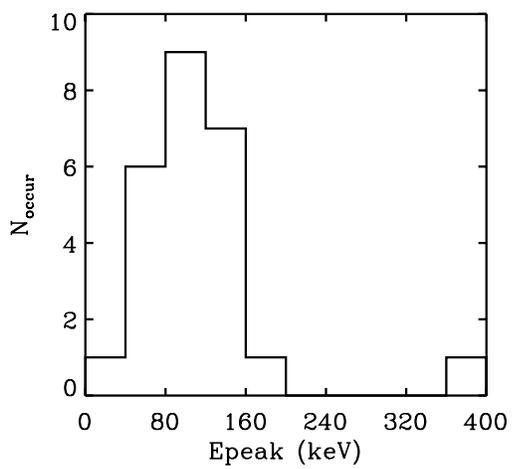

Fig11_Epeak-histo.eps

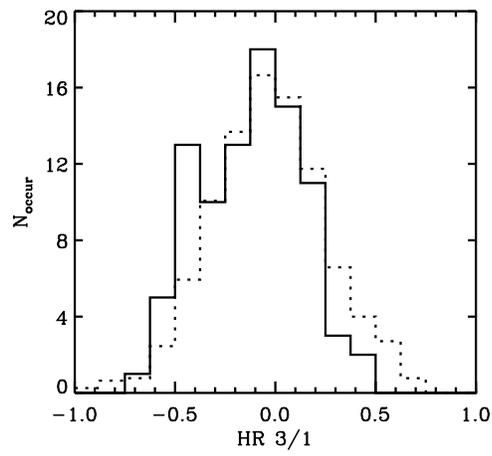

Fig12_HR31-histo.eps

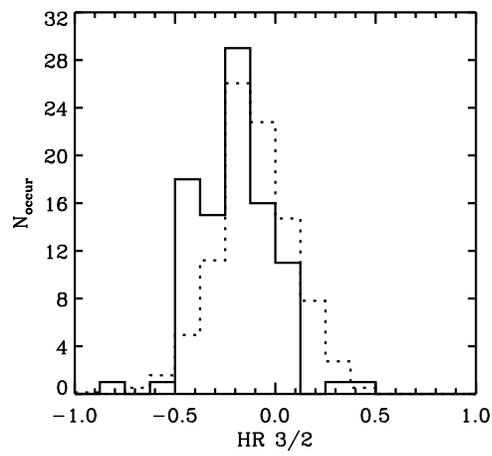

Fig13_HR32-histo.eps

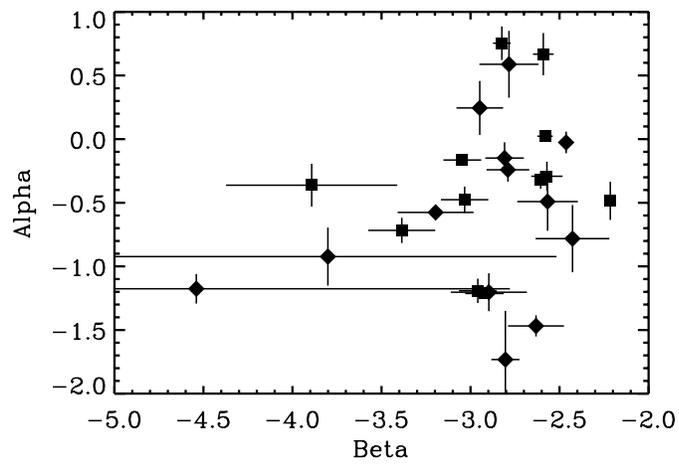

Fig14_Alpha-vs-Beta.eps

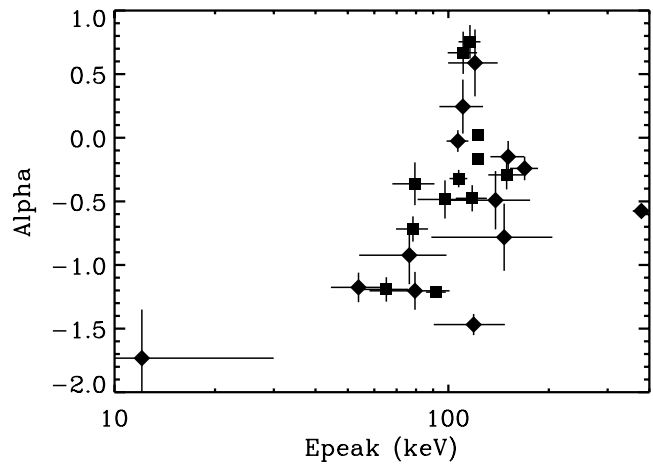

Fig15_Alpha-vs-Epeak.eps

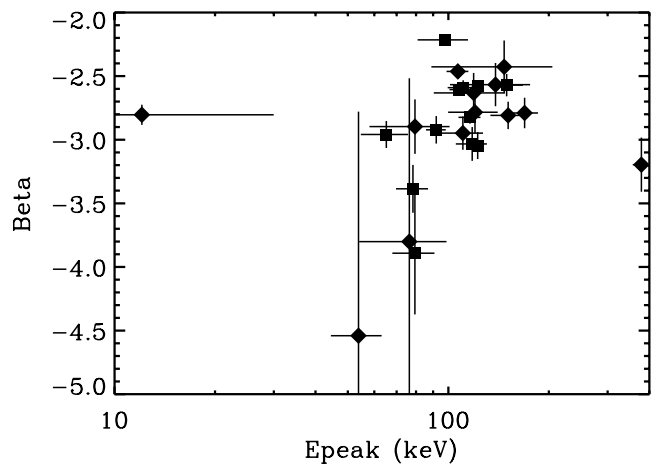

Fig16_Beta-vs-Epeak.eps

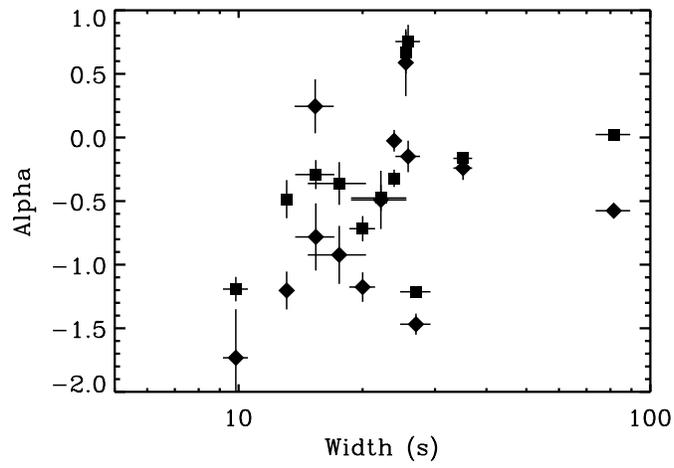

Fig17_Width-vs-Alpha.eps

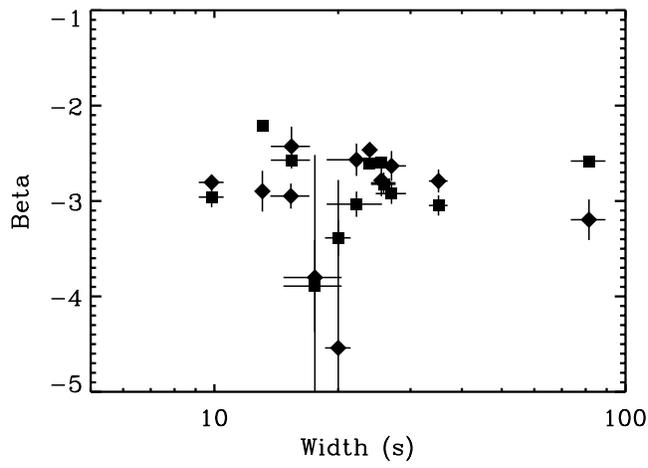

Fig18_Width-vs-Beta.eps

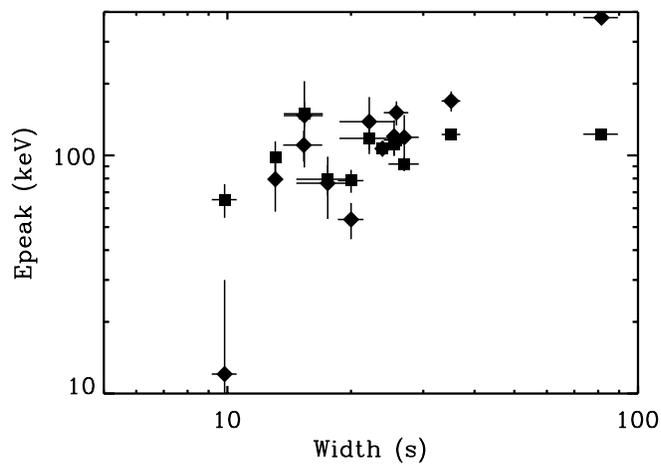

Fig19_Width-vs-Epeak.eps

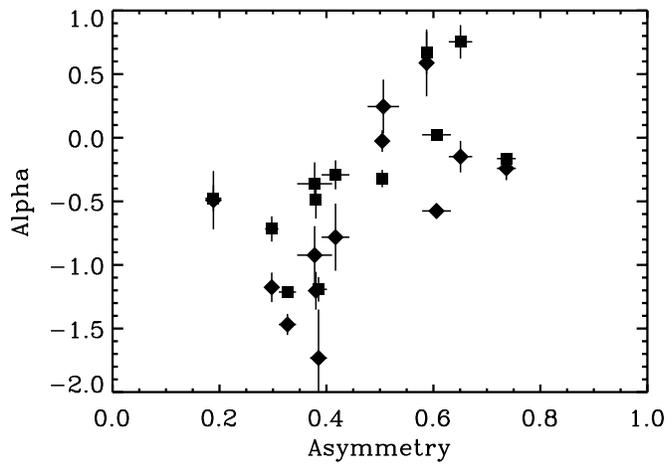

Fig20a_Asymmetry-vs-Alpha.eps

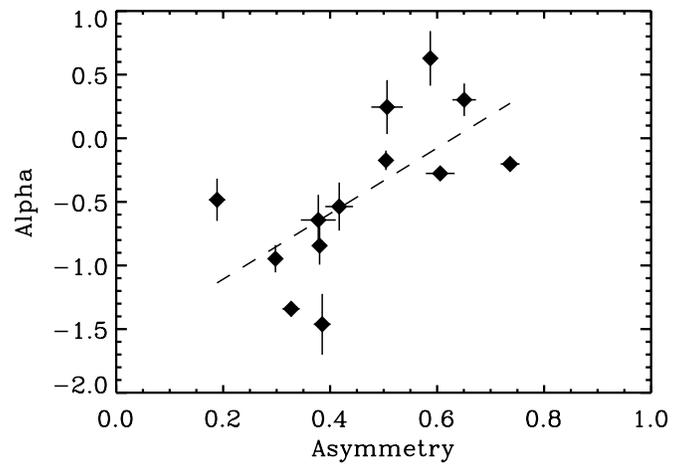

Fig20b_Asymmetry-vs-Alpha.avg.eps

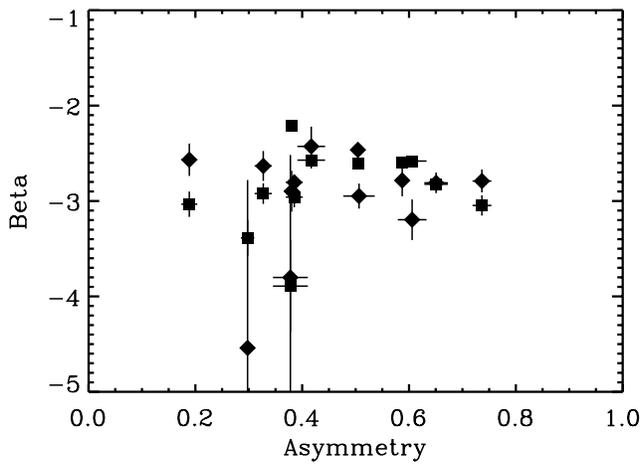

Fig21_Asymmetry-vs-Beta.eps

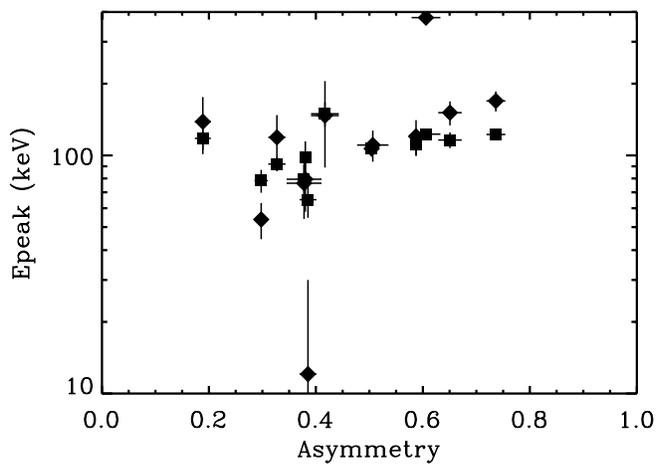

Fig22_Asymmetry-vs-Epeak.eps

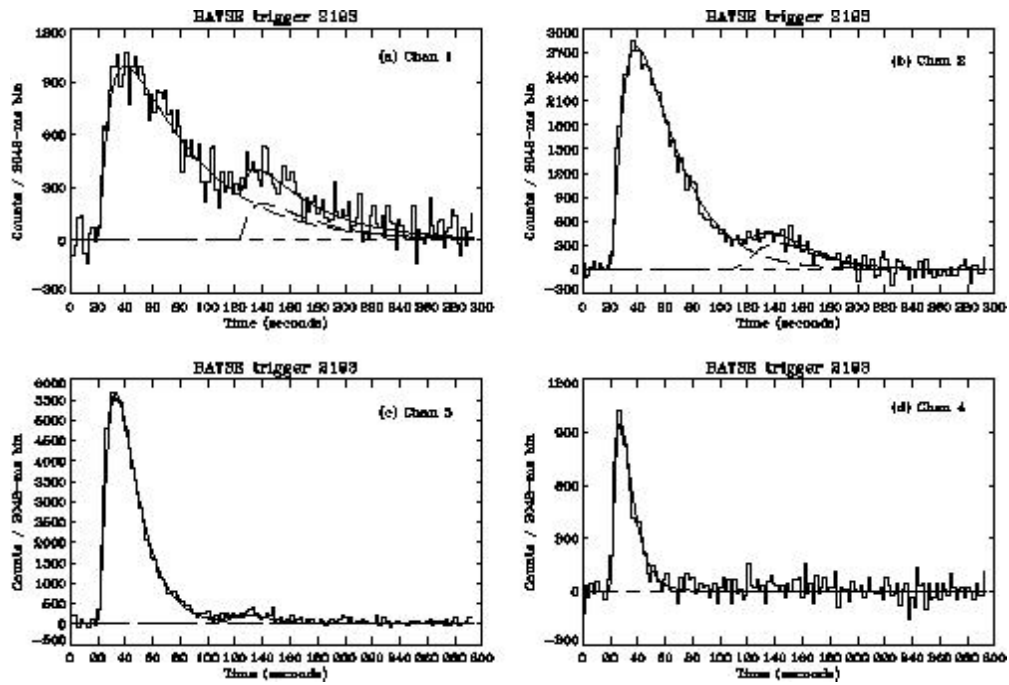

Fig_23a-d_2193_ch1.eps

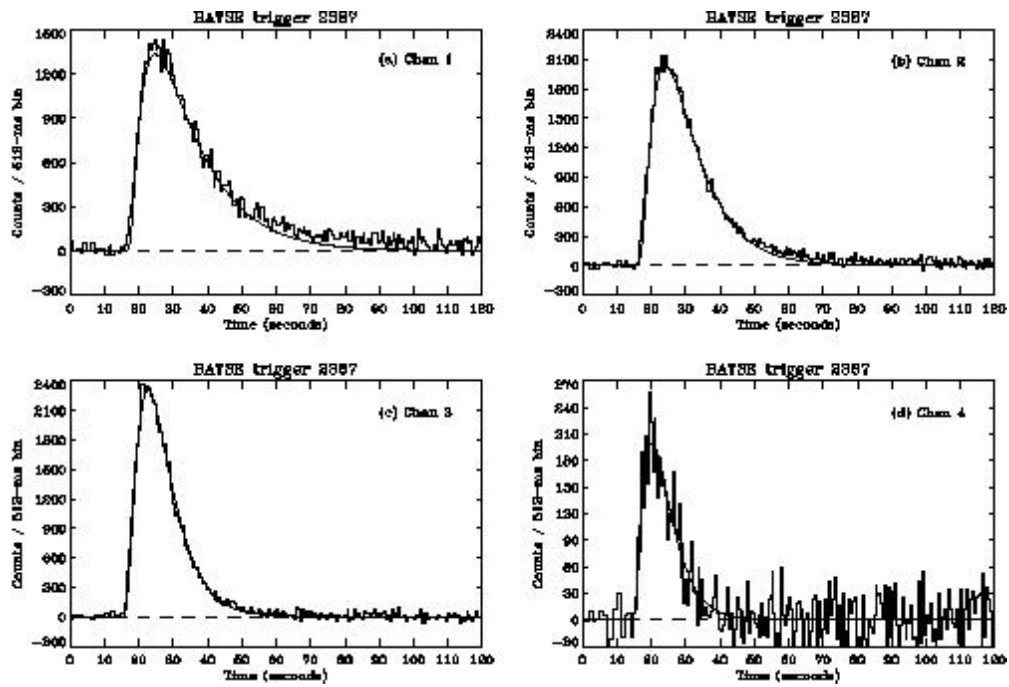

Fig_24a-d_2387_ch1.eps

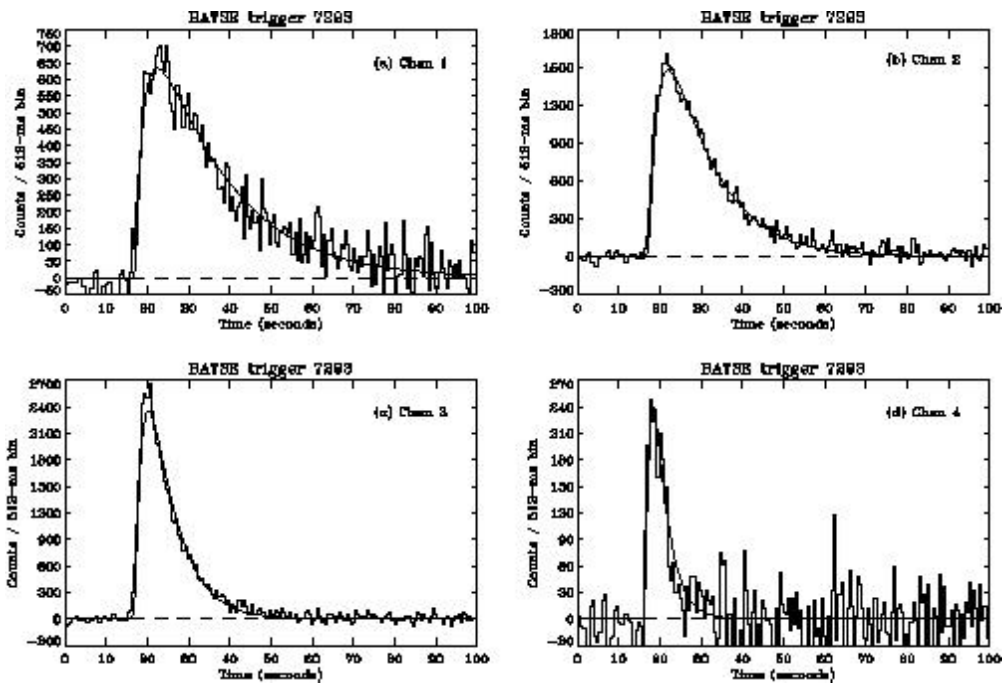

Fig_25a-d_7293_ch1.eps

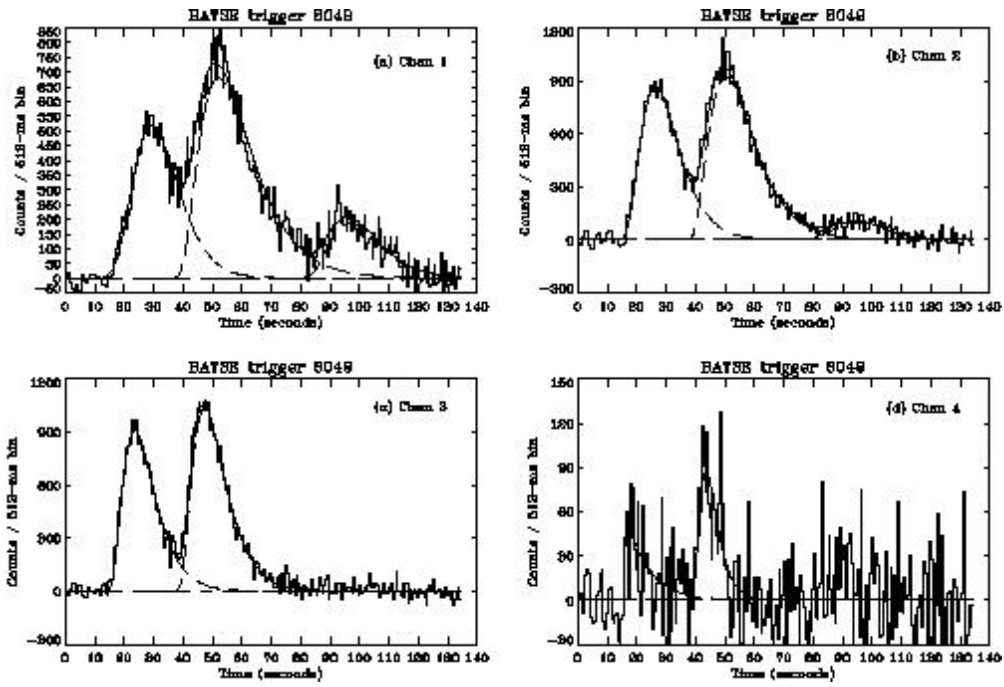

Fig_26a-d_8049_ch1.eps